\documentclass[twocolumn, journal]{IEEEtran}

\IEEEoverridecommandlockouts

\usepackage{float}
\usepackage{stfloats}
\usepackage{mathrsfs}
\usepackage{diagbox}
\usepackage{bm}
\usepackage{color}
\usepackage{graphicx}
\usepackage{amsmath}
\usepackage{amssymb}
\usepackage{algorithm}
\usepackage{algorithmic}

\usepackage{lineno}
\usepackage{lettrine}
\usepackage{subfigure}
\usepackage{epstopdf}
\usepackage{stfloats}
\usepackage{color}
\usepackage{graphicx}
\usepackage{amsmath}
\usepackage{amssymb}
\usepackage{algorithm}
\usepackage{algorithmic}
\usepackage{amsmath}
\usepackage{multirow}
\usepackage{booktabs}
\usepackage{array}
\usepackage{amsthm}
\usepackage{stfloats}
\usepackage{caption}
\usepackage{subfigure}
\usepackage{bm}
\usepackage{booktabs}
\usepackage{setspace}
\usepackage{makecell}
\usepackage{cases}

\newcommand{\be}{\begin{equation}}
	\newcommand{\ee}{\end{equation}}
\newcommand{\bea}{\begin{eqnarray}}
	\newcommand{\eea}{\end{eqnarray}}
\newcommand{\ba}{\begin{array}}
	\newcommand{\ea}{\end{array}}

\renewcommand{\thesubsubsection}{\thesubsection.\arabic{subsubsection}}

\newcounter{mytempeqncnt}
\allowdisplaybreaks[4]

\title{Joint Data Collection and Sensor Positioning in Multi-UAV-Assisted Wireless Sensor Network
	\thanks{Mingyue Zhu, Zhiqing Wei, Wangjun Jiang and Zhiyong Feng  are with School of Information and Communication Engineering, Beijing University of Posts and Telecommunications (BUPT), Beijing 100876, China. (e-mail: mingyue\_zhu@bupt.edu.cn, weizhiqing@bupt.edu.cn, jiangwangjun@bupt.edu.cn, fengzy@bupt.edu.cn).}
	\thanks{Chen Qiu is with Peng Cheng Laboratory, Shenzhen, China. (e-mail: qiuch@pcl.ac.cn).}
	\thanks{Huici Wu is with National Engineering Research Center of Mobile Network Technologies, Beijing University of Posts and Telecommunications (BUPT), Beijing, 100876, China, and also with Peng Cheng Laboratory, Shenzhen, China. (e-mail: dailywu@bupt.edu.cn).}
	}

\author{Mingyue Zhu,
	Zhiqing Wei,
	Chen Qiu, Wangjun Jiang,  Huici Wu, and Zhiying Feng
}
\begin{document}
\maketitle
\thispagestyle{empty}
\begin{abstract}
Due to the high mobility and easy deployment, unmanned aerial vehicles (UAVs) have attracted much attention in the field of wireless communication and positioning. To meet the challenges of lack
of infrastructure coverage, uncertain sensor position
and large amount of sensing data collection in wireless sensor network (WSN), this paper presents an efficient joint data collection and sensor positioning scheme for WSN supported by multiple UAVs. Specifically, a UAV is set as the main UAV to collect data, and other UAVs are used as auxiliary UAVs for sensor positioning using time difference of arrival (TDoA). A mixed-integer non-convex optimization problem with uncertain sensor position is established. The goal is to minimize the average positioning error of all sensors by jointly optimizing the UAV trajectories, sensor transmission schedule and positioning observation points (POPs). To solve this optimization model, the original problem is decomposed into two sub-problems based on the path discrete method. Firstly, the block coordinate descent (BCD) and successive convex approximation (SCA) techniques are applied to iteratively optimize the trajectory of the main UAV and the sensor transmission schedule, so as to maximize the minimum amount of data uploaded by the sensor. Then, based on the trajectory of the main UAV, a particle swarm optimization (PSO)-based algorithm is designed to optimize the POPs of UAVs. Finally, the spline curve is applied to generate the trajectories of auxiliary UAVs. The simulation results show that the proposed scheme can meet the requirements of data collection and has a good positioning performance. 	
\end{abstract}
	
	
\section{Introduction}
\label{sec:introduction}
Wireless sensor network (WSN) is an important component of the Internet of Things (IoT), which is composed of a large number of static or mobile sensors with sensing, computing and wireless communication capabilities. Due to the characteristics of wide coverage, low cost and remote monitoring, WSN is widely used in agriculture \cite{9350644}, industry \cite{9357334}, environmental monitoring \cite{9369359,7833202}, intelligent transportation \cite{9298484}, etc., which is responsible for sensing, collecting, processing and transmitting the sensing information \cite{9779853}. 

However, with the advancement and application of WSN, the amount of sensing data is also increasing significantly. However, data collection of sensors remains a challenge in the areas that lack infrastructure support, such as forests, oceans and islands. Besides, sensors are widely distributed and their positions are difficult to be acquired in advance \cite{9126800}. Therefore, it is necessary to locate sensors for WSN, which will facilitate identifying where events occur \cite{8502536,9196398}. In addition, the location information of sensor could be applied in the optimization of data collection scheme, further improving the performance of WSN via the collaboration
between data collection and sensor positioning. Hence, the
data collection and sensor positioning are crucial for WSN.

With the development of UAV technology, UAV with strong
coverage and mobility provides new opportunities for data
collection of WSN \cite{8417673,8468018,8641424}. In addition, the ability to be deployed quickly, move flexibly, and establish the line-of-sight
(LoS) connection with high probability makes UAV-assisted positioning highly attractive \cite{9991991}. The UAV equipped with
Global Positioning System (GPS) and wireless communication
modules can be used as mobile anchor nodes to estimate the
positions of sensors by broadcasting or receiving positioning
signals \cite{7438736}. Therefore, the joint data collection and sensor
positioning using UAVs is promising in WSN.

\subsection{Related Work}
Traditionally, sensor positioning and data collection in WSN
are addressed separately. For example, the data collection
schemes are designed assuming that the location information
of sensors is known \cite{9148990,8842600}. And the positioning methods
of sensors are addressed alone \cite{9748989}.

\subsubsection{UAV-assisted data collection schemes}
In terms of UAV-assisted data collection when the location information of sensors is known, Liu \emph{et al.} \cite{9286911} applied the age of information (AoI) to model the information timeliness of data collection. Then, a method of selecting data collection points is proposed. Dynamic programming and genetic algorithm (GA)
are applied to optimize the trajectory of UAV to minimize
the maximum and average AoI, respectively. To reduce the
energy consumption of WSN, Chen \emph{et al.} \cite{9749130} proposed a
data collection scheme considering network heterogeneity. A
method jointly optimizing the operation mode of sensors,
energy consumption threshold and cluster head selection is
proposed to extend the network lifetime. Chen \emph{et al.} \cite{9918630} studied the WSN with periodic data collection requirements. To minimize the flight trajectory of UAV and energy consumption of sensors, a two-stage task planning strategy
is proposed, which adopts ant colony optimization (ACO)
algorithm and data-aware strategy to optimize the trajectory
of UAV and the location of data collection points.

In terms of UAV-assisted data collection without the location
information of sensors, Wang \emph{et al.} \cite{8708930} proposed a heuristic algorithm to optimize the trajectory of UAV to minimize
the completion time of data transmission considering the
constraints of UAV flight speed, acceleration, communication
distance and communication throughput. In \cite{8737778}, Yin \emph{et al.} optimized the trajectory of UAV in the scenario of UAV-assisted cellular network to improve the uplink sum rate. With
the information of the position, transmit power and channel
state information of the ground user unknown, a deterministic
policy gradient algorithm is proposed to enable the
UAV to track the ground user.

\subsubsection{UAV-assisted sensor positioning schemes}
The
sensor positioning schemes in WSN include received signal
strength (RSS), time of arrival (ToA) and time difference of
arrival (TDoA) methods, which have high positioning accuracy
and reliability. Among them, RSS-based positioning methods
are widely applied in sensor positioning because of easy
deployment and low cost \cite{8125164,8647530,9536712}. In \cite{8125164}, a static sensor positioning scheme based on RSS is proposed, which takes into account the variation of path loss factor and shadow
factor due to UAV height, and explores the impact of the
number of UAVs and flight height on positioning error. On
this basis, Sallouha \emph{et al.} \cite{8647530} assumed that the UAV follows
a predetermined circular trajectory and collects RSS measurements at specific points. Then, the positioning accuracy is improved with limited energy by optimizing the UAV
trajectory parameters, including UAV height, hover time, flight
distance and the number of positioning points. Zou \emph{et al.} \cite{9536712} localized mobile nodes with multiple UAVs considering the time-varying of path-loss factor. The method combining least-mean-square (LMS) and iterative Newton gradient algorithm
is proposed. Since the RSS-based positioning schemes have 
low positioning accuracy, Yuan \emph{et al.} \cite{9815239} proposed a UAV path planning and ToA positioning scheme in the Non-LoS (NLoS)
environment. However, ToA is time-dependent and requires a
high-accurate time synchronization. In contrast, TDoA does
not require the time synchronization between UAV and sensor,
which is more practical compared with ToA scheme.

Overall, the above independent research on data collection
and sensor positioning has some deficiencies. In the scenario
of UAV-assisted data collection, some studies assume that the
location information of sensors is known without the design
of positioning scheme, which is not practical. Besides, if data
collection and sensor positioning are executed separately, additional communication and energy resources will be consumed.
Hence, the joint data collection and sensor positioning scheme
is required.

\subsubsection{UAV-assisted joint data collection and sensor positioning schemes} Wang \emph{et al.} \cite{8894454} applied a UAV to assist ground base stations (BSs) in data collection and positioning of devices. The device transmits data to the UAV or BS based on the transmission rate at each time slot. And the UAV only provide positioning services at positions with the best positioning performance on the trajectory. To minimize the maximum energy consumption of all devices and ensure the required positioning accuracy, a differential evolution (DE) is proposed to jointly optimize the UAV trajectory and device transmission schedule. However, the proposed solution requires the assistance of BSs in data collection and positioning, which cannot be applied to areas lacking infrastructure coverage. In addition, the population initialization of DE is a difficulty due to a large number of variables. In \cite{10039175}, Zhu \emph{et al.} applied a UAV to preliminarily realize joint sensor positioning and data collection through
RSS technology and hovering based on the designed trajectory.
Notwithstanding, the positioning accuracy is low due to RSS
technology and the energy consumption of UAV is high due
to the traversal trajectory.

\subsection{Contributions and Organization}
In view of the above limitations, we propose a joint data
collection and sensor positioning scheme using multiple UAVs
in WSN. A joint optimization framework is proposed to
minimize the average positioning error of all sensors while
satisfying the data collection requirements for sensors. The
main contributions of this paper are summarized as follows.

\begin{itemize}
	\item[1)] The multiple UAVs are applied as the data collector
	and mobile anchor to provide new solutions for data
	collection and sensor positioning. One UAV acts as the
	main UAV that is responsible for collecting sensing
	data, while the other UAVs apply TDoA technology in
	sensor positioning. Considering the constraints of the
	flight energy consumption, flight speed, requirements
	for data collection and communication range, a mixed-integer non-convex optimization model is established.
	The optimization model is to optimize the UAV flight
	trajectory, sensor transmission schedule and positioning
	observation points (POPs) jointly to minimize the average
	positioning error of sensors, while ensuring the amount of
	data collection. In particular, the optimization model takes
	into account the uncertainty of sensor position, which is
	more practical.
	
	\item[2)] The original problem is decomposed into two sub-problems, namely sub-problem 1: joint data collection
	and main UAV trajectory optimization, and sub-problem
	2: joint sensor positioning and auxiliary UAV trajectory
	optimization. For sub-problem 1, an iterative algorithm
	for solving mixed non-convex problems is proposed using the block coordinate descent (BCD) and successive convex approximation (SCA) techniques. Given the completion time of the task, the trajectory of main UAV and the sensor transmission schedule are iteratively optimized
	to maximize the minimum amount of data uploaded by
	sensors. The simulation results verify the feasibility of
	the proposed algorithm in different application scenarios.
	For sub-problem 2, based on the trajectory of main UAV
	optimized by sub-problem 1, a particle swarm optimization (PSO)-based algorithm is designed to optimize the
	POPs of UAV and minimize the average positioning error
	of sensors. Furthermore, the spline curve is applied to
	generate the trajectory of auxiliary UAV. The simulation
	results show that the proposed scheme can achieve the
	meter-level positioning accuracy with only three UAVs.	
\end{itemize}

The rest of this paper is organized as follows. Section II
introduces the system model and the problem formulation.
In Section III, we propose an efficient optimization method
to solve the mixed-integer non-convex optimization problem.
In Section IV, the numerical simulation results are provided
to verify the proposed scheme and evaluate its performance.
Section V summarizes this paper.

\section{System Model and Problem Formulation}

Consider a WSN with an area of $L \times L  ($m$^2)$. There are $M$ sensors  randomly deployed in fixed positions to collect data from their surrounding environment, where the set of sensors is denoted by ${\cal M} = \left\{ {1, \ldots ,m, \ldots ,M} \right\}$. Assume that
the possible deployment space of sensors can be obtained
from prior information. The real coordinates of sensors can be
defined as ${\left( {{\bf{s}}_m^T,{h_m}} \right)^T}$, with $h_m$ being the vertical coordinate and
\begin{equation}
	{{\bf{s}}_m} = {\left( {x_m^s,y_m^s} \right)^T} \in {\mathbb{R}^{2 \times 1}}
	\label{equ_01}
\end{equation}
being the horizontal coordinate. Due to the portability and
hovering ability, rotorcraft UAVs are adopted in this paper. Let ${\cal N} = \left\{ {1, \ldots ,n, \ldots ,N} \right\}$ be the set of UAVs, including one main UAV and $N-1$ auxiliary UAVs. Assume that UAVs start at $t = 0$ and fly at a fixed altitude $H$. The period for positioning and data collection is $T$. The horizontal coordinates of UAVs at $t \in \left[ {0,T} \right]$ can be expressed as
\begin{equation}
	{{\bf{u}}_n}\left( t \right) = {\left( {x_n^u\left( t \right),y_n^u\left( t \right)} \right)^T} \in {\mathbb{R}^{2 \times 1}}.
	\label{equ_02}
\end{equation}

The speed of UAVs can be denoted by the time-derivative of ${{\bf{u}}_n}\left( t \right)$
\begin{equation}
	{{\bf{v}}_n}\left( t \right) \buildrel \Delta \over = {{\bf{\dot u}}_n}\left( t \right),
	\label{equ_03}
\end{equation}
where ${{\bf{\dot u}}_n}\left( t \right)$ is the derivative operation. To avoid unnecessary energy consumption, the sensor is
generally in a sleep state. After receiving the activation signal
from the UAV 1, the sensor will transmit the collected data
to UAVs in the way of backscattering \cite{8752043}. The task period
is equally divided into $W$ time slots of length $\tau$ with ${\cal W} = \left\{ {1, \ldots ,w, \ldots ,W} \right\}$ being the index. The position and speed of UAVs and the channel environment between UAVs and
sensors are stable in each time slot $\tau$ \cite{8894454}. Based on path discrete method, the continuous trajectory of UAVs is divided into trajectory sequences. In the $w$-th time slot, the position and speed of the UAV can be expressed as ${{\bf{u}}_n}\left[ w \right]$ and ${{\bf{v}}_n}\left[ w \right]$, respectively.

\begin{figure}[t]
	\centering
	\includegraphics[width=0.3\textheight]{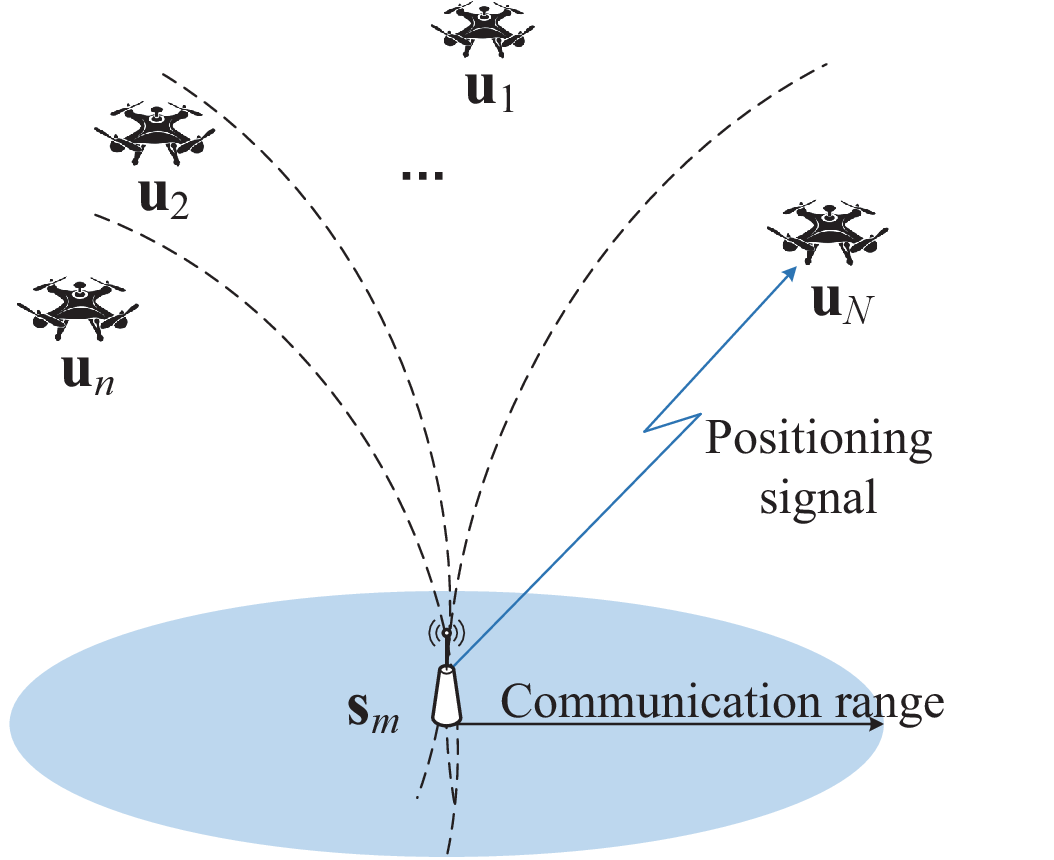}
	\DeclareGraphicsExtensions.
	\caption{TDoA positioning assisted by $N$ UAVs.}
	\label{fig 1}
\end{figure}

\subsection{Model of TDoA Localization}
The positioning of sensors in WSN is usually implemented
by using the TDoA algorithm, which has the advantages of low
complexity and low time synchronization requirements \cite{8732484}. Suppose that the synchronization accuracy between UAVs can
meet the requirements of TDoA. As shown in \textcolor[rgb]{0,0.4471,0.4039}{Fig. \ref{fig 1}}, there
are $N$ UAVs to locate the sensor, where UAV 1 is used as
the main UAV, and others are used as auxiliary UAVs. The
actual TDoA of the positioning signal received by multiple
time-synchronized UAVs is 
\begin{equation}
	t_{n1}^0 = t_n^0 - t_1^0,{\kern 1pt} {\kern 1pt} {\kern 1pt} {\kern 1pt} {\kern 1pt} {\kern 1pt} {\kern 1pt} {\kern 1pt} {\kern 1pt} {\kern 1pt} {\kern 1pt} {\kern 1pt} {\kern 1pt} {\kern 1pt} {\kern 1pt} {\kern 1pt} {\kern 1pt} {\kern 1pt} i = 2,3, \ldots ,N,
	\label{equ 1}
\end{equation}
where $t_n^0$ is the ToA of the positioning signal from the sensor to the $n$-th UAV. The estimated location of the sensor can
be obtained by solving the hyperbolic equations composed of $N-1$ observations below.
\begin{equation}
	\begin{aligned}
		ct_{n1}^0 &= {r_n} - {r_1}= \sqrt {{{\left\| {{{\bf{u}}_n}\left[ w \right] - {{\bf{s}}_m}} \right\|}^2} + {{\left( {H - {h_m}} \right)}^2}} \\  &- \sqrt {{{\left\| {{{\bf{u}}_1}\left[ w \right] - {{\bf{s}}_m}} \right\|}^2} + {{\left( {H - {h_m}} \right)}^2}} ,{\kern 1pt} {\kern 1pt} {\kern 1pt} {\kern 1pt} {\kern 1pt} {\kern 1pt} {\kern 1pt} {\kern 1pt} {\kern 1pt} {\kern 1pt} {\kern 1pt} {\kern 1pt}{\kern 1pt} i = 2,3, \ldots ,N,
		\label{equ 2}
	\end{aligned}
\end{equation}
where $c$ is the speed of electromagnetic wave. As  \textcolor[rgb]{0,0.4471,0.4039}{Fig. \ref{fig 1}} shows, the estimated position of the sensor is the intersection
of multiple hyperbolas. At least three UAVs are required for
two-dimensional (2D) positioning. Considering the random
noise, the matrix of TDoA localization model can be expressed
as
\begin{equation}
	\begin{aligned}
		{\bf{r}} &= {{\bf{r}}^0} + {\bf{e}} \\
		&= {\left[ {ct_{21}^0,ct_{31}^0, \cdots ,ct_{N1}^0} \right]^T} + {\left[ {{e_{21}},{e_{31}}, \cdots ,{e_{N1}}} \right]^T},
		\label{equ 3}
	\end{aligned}
\end{equation}
where ${e_{n1}} = c\Delta {t_{n1}}$ is the measurement error of the TDoA,
and follows the zero mean Gaussian distribution with variance $\delta ^2$. According to \cite{5937261}, the Cramer-Rao lower bound (CRLB) based on TDoA method is
\begin{equation}
	{\rm{CRLB}}\left( {\bf{s}} \right) = {c^2} \textup {trace}\left( {{{\left( {{\bf{HQ}}_t^{ - 1}{{\bf{H}}^T}} \right)}^{ - 1}}} \right){\kern 1pt} {\kern 1pt} {\kern 1pt},
	\label{equ 4}
\end{equation}
where ${{\bf{Q}}_t} = {\rm{E}}\left\{ {{\bf{e}}{{\bf{e}}^T}} \right\}$ is the covariance matrix of the observed values of the TDoA. ${\bf{H}}$ is the Jacobian matrix of the equation, which is expressed as
\begin{equation}
	{\bf{H}} = \left[ {\begin{array}{*{20}{c}}
			{\frac{{\partial ct_{21}^0}}{{\partial x}}}& \cdots &{\frac{{\partial ct_{N1}^0}}{{\partial x}}}\\
			{\frac{{\partial ct_{21}^0}}{{\partial y}}}& \cdots &{\frac{{\partial ct_{N1}^0}}{{\partial y}}}
	\end{array}} \right]{\kern 1pt},
	\label{equ 5}
\end{equation}

The partial derivative ${{\partial ct_{i1}^0} \mathord{\left/
		{\vphantom {{\partial ct_{i1}^0} {\partial x}}} \right.
		\kern-\nulldelimiterspace} {\partial x}}$ and ${{\partial ct_{i1}^0} \mathord{\left/
		{\vphantom {{\partial ct_{i1}^0} {\partial y}}} \right.
		\kern-\nulldelimiterspace} {\partial y}}$ can be expressed as
\begin{equation}
	\frac{{\partial ct_{21}^0}}{{\partial x}} = \frac{{x_1^u - x_m^s}}{{{r_1}}} - \frac{{x_2^u - x_m^s}}{{{r_2}}},
	\label{equ 6}
\end{equation} 

\begin{equation}
	\frac{{\partial ct_{21}^0}}{{\partial y}} = \frac{{y_1^u - y_m^s}}{{{r_1}}} - \frac{{y_2^u - y_m^s}}{{{r_2}}}.
	\label{equ 61}
\end{equation} 

Obviously, the positioning performance of the sensor assisted
by UAVs is directly related to the position of UAVs. It is
assumed that UAVs only select the position with the best
performance as the POPs in their flight trajectory. Therefore,
the trajectory design of UAVs is very important.

\subsection{Model of Ground-to-Air Channel}

Due to the flying altitude of the UAV, the communication
between the UAV and the sensor is more likely to be dominated by the LoS link, especially in the environment
with few obstacles \cite{9701330,Qualcomm}. The impact of small-scale fading on signal attenuation is usually ignored in the scenario of UAV-assisted data collection \cite{8757098,8779596}. Therefore, we only consider the path loss in the ground-to-air (G2A) channel, and the channel gain from sensors to UAVs can
be expressed as \cite{8743402}
\begin{equation}
	\begin{aligned}
		P{L_m}\left[ w \right] &= {\beta _0}d_{_m}^{ - \alpha }\left[ w \right] \\ 
		&= {\beta _0}{\left( {{{\left\| {{{\bf{u}}_1}\left[ w \right] - {{\bf{s}}_m}} \right\|}^2} + {{\left( {H - {h_m}} \right)}^2}} \right)^{{{ - \alpha } \mathord{\left/
						{\vphantom {{ - \alpha } 2}} \right.
						\kern-\nulldelimiterspace} 2}}},
		\label{equ 7}
	\end{aligned}
\end{equation} 
where $\beta _0$ is the free space path loss at the reference distance of 1 m, and $\alpha  \ge 2$ is the path loss factor. ${{\bf{u}}_1}\left[ w \right]$ and ${{\bf{s}}_m}$ represent the position of UAV $1$ and sensor $m$ in the $w$-th time slot, respectively.

Specifically, the UAV 1 can establish communication links
with at most ${K_{\max }}\left( {1 \le {K_{\max }} \le M} \right)$ sensors and receive data in each time slot. The bandwidth $B_0$ is evenly divided into $K$ sensors to upload data at the same time, and the set of connected sensors denoted by ${\cal K}$. Then the bandwidth allocated to each sensor is
\begin{equation}
	B\left[ w \right] = \frac{{{B_0}}}{K}.
	\label{equ 8}
\end{equation} 

The signal to interference plus noise ratio (SINR) between the sensor $m$ and UAV $1$ can be expressed as
\begin{equation}
	{\gamma _m}\left[ w \right] = \frac{{{P_t} \cdot P{L_m}\left[ w \right]}}{{\sum\limits_{k \in {\cal K}\backslash m} {{P_t} \cdot P{L_m}\left[ w \right]}  + {\sigma ^2}}},
	\label{equ 9}
\end{equation} 
where ${P_t}$ is the fixed transmission power of the sensor, $\sum\limits_{k \in {\cal K}\backslash m} {{P_t} \cdot P{L_m}\left[ w \right]} $ is the co-channel interference, and ${\sigma ^2}$ is the  power of additive Gaussian white noise (AWGN). If the UAV $1$ serves the sensor in the $w$-th time slot, the transmission rate (bits/s) can be expressed as \cite{9696083,9943536}
\begin{equation}
	{R_m}\left[ w \right]{\rm{ = }}B\left[ w \right]{\log _2}\left( {1 + {\gamma _m}\left[ w \right]} \right).
	\label{equ 10}
\end{equation} 

\begin{figure}[t]
	\centering
	\includegraphics[width=0.38\textheight]{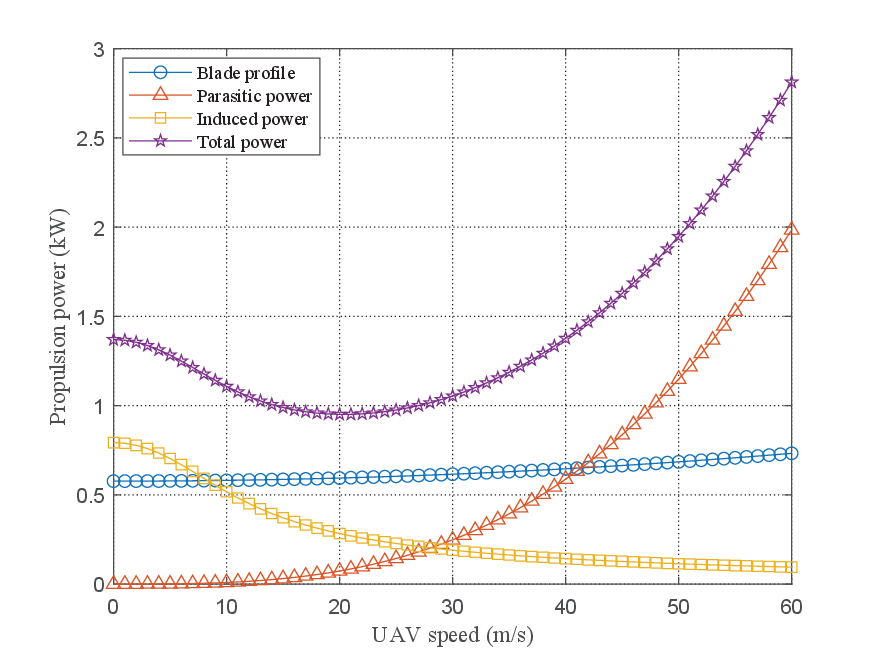}
	\DeclareGraphicsExtensions.
	\caption{Propulsion powers versus UAV speed.}
	\label{fig 2}
\end{figure}

The wake-up communication scheduling method is adopted
in this paper \cite{8779596,9013148}. The schedule of the transmission can be controlled to determine the sensor to communicate in each
time slot. Assume that the UAV 1 can only collect data from
one sensor in a time slot, i.e., ${K_{\max }} = 1$. Define a binary variable ${x_m}\left[ w \right],\forall m,w$ to represent data transmission in WSN. ${x_m}\left[ w \right] = 1$ denotes that sensor $m$ transmits data to UAV $1$ in the $w$-th time slot, while ${x_m}\left[ w \right] = 0$ denotes the
reverse process. Therefore, we have the following transmission
schedule constraints.
\begin{equation}
	{x_m}\left[ w \right] \in \left\{ {0,1} \right\}{\kern 1pt} {\kern 1pt} {\kern 1pt} {\kern 1pt}  {\kern 1pt} {\kern 1pt} {\kern 1pt} \forall m \in {\cal M},w \in {\cal W},
	\label{equ 11}
\end{equation}
\begin{equation}
	\sum\limits_{m = 1}^M {{x_m}\left[ w \right]}  \le 1{\kern 1pt} {\kern 1pt} {\kern 1pt} {\kern 1pt} {\kern 1pt} {\kern 1pt} {\kern 1pt} {\kern 1pt} {\kern 1pt} {\kern 1pt} {\kern 1pt} {\kern 1pt} {\kern 1pt}  \forall w \in {\cal W}.
	\label{equ 12}
\end{equation}

Furthermore, the amount of data uploaded by sensor $m$ in the whole task is
\begin{equation}
	{I_m} = \tau \sum\limits_{w = 1}^W {{R_m}\left[ w \right] \cdot {x_m}\left[ w \right]} {\kern 1pt} {\kern 1pt} {\kern 1pt} {\kern 1pt} {\kern 1pt} {\kern 1pt} {\kern 1pt} {\kern 1pt} {\kern 1pt} {\kern 1pt} \forall m \in {\cal M}.
	\label{equ 13}
\end{equation}

According to \eqref{equ 10} and \eqref{equ 13}, it is revealed  that the data
transmission rate is directly related to the distance between
the UAV and the sensor, namely, the instantaneous position
of UAV 1 when collecting data will affect the efficiency of
data collection. Therefore, it is necessary to design reasonable transmission schedule and UAV trajectory to ensure the
performance of data collection.

\begin{algorithm}[!t]
	\caption{BCD for sub-problem 1}
	\begin{algorithmic}[1]
		\STATE \textbf{Initialize} $\left\{ {{\bf{u}}_1^0\left[ w \right]} \right\}$, ${{\bf{X}}^0}$ and let $l \leftarrow 0$.
		\STATE \textbf{repeat}
		\STATE For given $\left\{ {{\bf{u}}_1^l\left[ w \right]} \right\}$, solve \textup(sP1-1) and denote the optimal solution as ${{\bf{X}}^{l + 1}}$.
		\STATE For given ${{\bf{X}}^{l + 1}}$, solve \textup(sP1-4) and denote the optimal solution as $\left\{ {{\bf{u}}_1^{l+1}\left[ w \right]} \right\}$.
		\STATE $l \leftarrow l + 1$.
		\STATE \textbf{until} $l = {l_{\max }}$.
	\end{algorithmic}
\end{algorithm}

\subsection{Model of UAV Energy Consumption}
The energy consumption of UAV is mainly divided into
two parts: communication energy consumption (signal transmission, reception and processing) and propulsion energy
consumption (flying and hovering). Since the propulsion energy consumption is usually much larger than communication
energy consumption, the communication energy consumption
can be ignored \cite{7888557}. According to \cite{8647595}, when the UAV flies at speed ${{\bf{v}}_n}\left[ w \right]$, the propulsion energy consumption has three components, which can be modeled as 
\begin{equation}
	\begin{aligned}
		P\left( {{{\bf{v}}_n}\left[ w \right]} \right) &= \underbrace {{k_b}\left( {1 + \frac{{3{{\left\| {{{\bf{v}}_n}\left[ w \right]} \right\|}^2}}}{{v_t^2}}} \right)}_{\textup{Blade profile}} + \underbrace {\frac{1}{2}\rho s{d_r}A{{\left\| {{{\bf{v}}_n}\left[ w \right]} \right\|}^3}}_{\textup{Parasitic power}} \\
		&+ \underbrace {{k_i}{{\left( {\sqrt {1 + \frac{{{{\left\| {{{\bf{v}}_n}\left[ w \right]} \right\|}^4}}}{{4v_0^4}}}  - \frac{{{{\left\| {{{\bf{v}}_n}\left[ w \right]} \right\|}^2}}}{{2v_0^2}}} \right)}^{{1 \mathord{\left/
							{\vphantom {1 2}} \right.
							\kern-\nulldelimiterspace} 2}}}}_{\textup{Induced power}},
		\label{equ 14}
	\end{aligned}
\end{equation}
where $k_b$ and $k_i$ are the blade profile power and induced power when the UAV hovers respectively. $\rho $ is the air density, $s$ is the rotor solidity, $d_r$ is the drag coefficient of UAV and $A$ is the rotor surface area. $v_t$ and $v_0$ are the blade tip speed and mean rotor induced speed in hovering respectively. When ${{\bf{v}}_n}\left[ w \right] = {{\bf{0}}^T}$, the hovering power of UAV can be obtained as ${P_h} = {k_b} + {k_i}$. \textcolor[rgb]{0,0.4471,0.4039}{Fig. \ref{fig 2}} shows the curve of total power and three power components as the rate changes.

\subsection{Problem Formulation}
Define ${\bf{U}} = \left\{ {{{\bf{u}}_n}\left[ w \right],\forall n,w} \right\}$ and ${\bf{X}} = \left\{ {{x_m}\left[ w \right],\forall m,w} \right\}$. To jointly optimize the UAV trajectory $\bf{U}$ and sensor transmission scheduling $\bf{X}$, we need to minimize the average positioning error of sensors, while ensuring the successful data uploading. Then, the optimization problem can be formulated as
\begin{align}
	\textup{(P0)}:{\kern 1pt} {\kern 1pt} &\mathop {\min }\limits_{{\bf{X}},{\bf{U}}}{\kern 1pt} {\kern 1pt} {\kern 1pt}{\kern 1pt} \frac{{\sum\limits_{m = 1}^M {{\rm{CRL}}{{\rm{B}}_m}\left( {{{\bf{u}}_1}\left[ w \right], \ldots ,{{\bf{u}}_n}\left[ w \right]} \right)} }}{M} \label{YY}\\
	\mbox{s.t.}\quad 	
	&\tau \sum\limits_{w = 1}^W {P\left( {\left\| {{{\bf{v}}_n}\left[ w \right]} \right\|} \right)}  \le {E_{\max }}, {\kern 1pt} {\kern 1pt} {\kern 1pt} {\kern 1pt} {\kern 1pt} \forall n \in {\cal N}, \tag{\ref{YY}{a}} \label{YYa}\\
	&{x_m}\left[ w \right] \in \left\{ {0,1} \right\},{\kern 1pt} {\kern 1pt} {\kern 1pt} {\kern 1pt} {\kern 1pt}\forall m \in {\cal M},w \in {\cal W}, \tag{\ref{YY}{b}} \label{YYb}\\
	&\sum\limits_{m = 1}^M {{x_m}\left[ w \right]}  \le 1,{\kern 1pt} {\kern 1pt} {\kern 1pt} {\kern 1pt} {\kern 1pt}\forall w \in {\cal W}, \tag{\ref{YY}{c}} \label{YYc}\\
	&\tau \sum\limits_{w = 1}^W {{R_m}\left[ w \right] \cdot {x_m}\left[ w \right]}  \ge {I_{\min }},{\kern 1pt} {\kern 1pt} {\kern 1pt} {\kern 1pt} {\kern 1pt}\forall m \in {\cal M}, \tag{\ref{YY}{d}} \label{YYd}\\
	&\left\| {{{\bf{u}}_n}\left[ w \right] - {{\bf{u}}_n}\left[ {w - 1} \right]} \right\| \le \tau {V_{\max }},{\kern 1pt} \forall n \in {\cal N},w \in {\cal W}, \tag{\ref{YY}{e}} \label{YYe}\\
	&\left\| {{{\bf{u}}_i}\left[ w \right] - {{\bf{u}}_1}\left[ w \right]} \right\| \le {R_{\max }},{\kern 1pt} {\kern 1pt} {\kern 1pt} {\kern 1pt} {\kern 1pt} \forall w \in {\cal W},{\kern 1pt} {\kern 1pt} i \ne 1, \tag{\ref{YY}{f}} \label{YYf}\\
	&{{\bf{u}}_n}\left[ 0 \right] = {{\bf{u}}_s},{{\bf{u}}_n}\left[ W \right] = {{\bf{u}}_e},{\kern 1pt} {\kern 1pt} {\kern 1pt} {\kern 1pt} {\kern 1pt}\forall n \in {\cal N}, \tag{\ref{YY}{g}} \label{YYg}
\end{align}
where ${V_{\max }}$ is the maximum speed of UAVs, ${R_{\max }}$ is the maximum communication distance between  the auxiliary UAV and UAV $1$. \eqref{YYa} is the constraint of maximum flight energy consumption of UAVs. \eqref{YYb} limits the feasible range of optimization variable $\bf{X}$, and \eqref{YYc} ensures that the number of sensors accessing UAV $1$ at the same time does not exceed $1$. \eqref{YYd} ensures that the amount of data uploaded by each sensor can meet the requirements. \eqref{YYe} is the constraint of maximum flight speed. \eqref{YYf}  represents the communication performance limitations of UAVs respectively, which ensures that the auxiliary UAV can maintain communication with UAV $1$ during positioning. \eqref{YYg} represents the start and end points of UAV trajectories. The UAV cannot collect a large amount of data with extremely low energy. In practical applications, the values of ${E_{\max}}$ and ${I_{\min}}$ can be continuously adjusted to ensure that at least one solution simultaneously satisfies all constraints \cite{8714077}.

Note that the calculation of data transmission rate and CRLB both requires the accurate position of the sensor in problem \textup{(P0)}. Therefore,
we build the uncertainty model of the sensor by making
a circle with rough position as the center and uncertainty
parameter as the radius \cite{8894454}. By assuming that the sensor is
located at the farthest distance from UAV 1 in the model, each
sensor can upload enough data when the data transmission rate
is the lowest. In the same way, when estimating the location
of the sensor, multiple sampling points are scattered in the
uncertainty model to calculate the CRLB, and the points with
maximum value are selected as the POPs.

This model is also suitable for scenarios with optimization requirements for flight altitude of UAVs. Some constraints on the vertical trajectory need to be added, i.e. \cite{9121338,9875063},
\begin{equation}
	\left\| {{H_n}\left[ w \right] - {H_n}\left[ {w - 1} \right]} \right\| \le \tau {V_{z - \max }},{\kern 1pt} \forall n \in {\cal N},w \in {\cal W},
	\label{equ 113}
\end{equation}
\begin{equation}
	{H_n}\left[ 0 \right] = {H_s},{H_n}\left[ W \right] = {H_e},\forall n \in {\cal N},
	\label{equ 114}
\end{equation}
\begin{equation}
	{H_{\min }} \le {H_n}\left[ w \right] \le {H_{\max }},\forall n \in {\cal N},w \in {\cal W},
	\label{equ 115}
\end{equation}
where \eqref{equ 113} is the constraint of the vertical flight speed. \eqref{equ 114} and \eqref{equ 115} constrain the flight height of UAVs within each time slot. Besides, considering the three-dimensional (3D) trajectory, the constraint \eqref{YYf} needs to be rewritten as
\begin{equation}
	\begin{aligned}
		&{\left\| {{{\bf{u}}_i}\left[ w \right] - {{\bf{u}}_1}\left[ w \right]} \right\|^2} + {\left\| {{H_n}\left[ w \right] - {H_n}\left[ {w - 1} \right]} \right\|^2} \\
		&\le R_{\max }^2,{\kern 1pt} \forall w \in {\cal W},{\kern 1pt} {\kern 1pt} i \ne 1.
		\label{equ 116}
	\end{aligned}
\end{equation}

However, the modification of (19a) is the main challenge in building and solving the 3D model. On the one hand, in 3D scenarios, the model of energy consumption needs to consider more factors, such as the changing lift and drag of the UAV \cite{2021Path,2019Energy,8647595}. On the other hand, more complex formulas result in a change in the problem structure, which means that the optimization problem is difficult to convert into a convex problem to solve.

\begin{figure}[t]
	\centering
	\includegraphics[width=0.36\textheight]{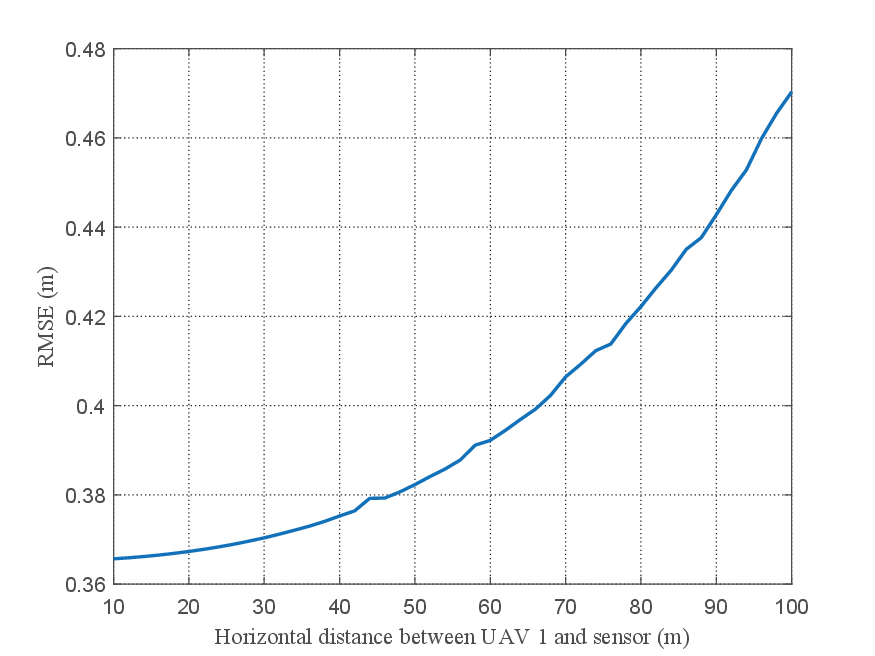}
	\DeclareGraphicsExtensions.
	\caption{RMSE versus the distance between UAV 1 and the sensor.}
	\label{fig 3}
\end{figure}

\section{Proposed optimization method}
The optimization problem \textup{(P0)} involves infinite variables,
binary constraints and non-convex constraints, and the objective function is not a closed-form expression. Therefore, \textup{(P0)}
is a mixed-integer non-convex optimization problem, which
is difficult to find the optimal solution directly. According
to the analysis in Section II, the transmission rate in data
collection is affected by the distance between the UAV 1
and the sensor, while the CRLB for sensor positioning is
related to the relative position of all UAVs and the sensor.
Therefore, we divide original problem \textup{(P0)} into two sub-problems. In sub-problem \textup{(sP1)}, we optimize the trajectory of
UAV 1 and the transmission schedule of sensors to maximize
the amount of data uploaded. In sub-problem \textup{(sP2)}, based
on the trajectory points of UAV 1, a PSO-based optimization
method is designed to determine the optimal POPs of UAVs
and minimize the average CRLB. Furthermore, the trajectories
of auxiliary UAVs are generated by spline curve. The solution obtained using the above is approximately optimal, but has lower complexity.

\subsection{\textup{(sP1)}: Transmission Scheduling and UAV 1 Trajectory Optimization}
For sub-problem \textup{(sP1)}, we apply alternating iteration technique to optimize the transmission scheduling of sensors and the trajectory of UAV $1$, i.e., ${\bf{X}}$ and ${{\bf{u}}_1}\left[ w \right]$. The details of BCD method are summarized in Algorithm 1. Sub-problem \textup{(sP1)} can be formulated as
\begin{align}
	\textup{(sP1)}:{\kern 1pt} {\kern 1pt} &\mathop {\max }\limits_{\lambda ,{\bf{X}},\left\{ {{{\bf{u}}_1}\left[ w \right]} \right\}} \lambda \label{XX}\\
	\mbox{s.t.}\quad 	
	&\tau \sum\limits_{w = 1}^W {P\left( {\left\| {{{\bf{v}}_1}\left[ w \right]} \right\|} \right)}  \le {E_{\max }}, \tag{\ref{XX}{a}} \label{XXa}\\
	&{x_m}\left[ w \right] \in \left\{ {0,1} \right\},\,\, \forall m \in {\cal M},w \in {\cal W}, \tag{\ref{XX}{b}} \label{XXb}\\
	&\sum\limits_{m = 1}^M {{x_m}\left[ w \right]}  \le 1,\,\, \forall w \in {\cal W}, \tag{\ref{XX}{c}} \label{XXc}\\
	&\frac{\tau }{{{I_{\min }}}}\sum\limits_{w = 1}^W {{R_m}\left[ w \right] \cdot {x_m}\left[ w \right]}  \ge \lambda,\,\, \forall m \in {\cal M}, \tag{\ref{XX}{d}} \label{XXd}\\
	&\left\| {{{\bf{u}}_1}\left[ w \right] - {{\bf{u}}_1}\left[ {w - 1} \right]} \right\| \le \tau {V_{\max }},\forall n \in {\cal N},w \in {\cal W}, \tag{\ref{XX}{e}} \label{XXe}\\
	&{{\bf{u}}_1}\left[ 0 \right] = {{\bf{u}}_s},{{\bf{u}}_1}\left[ W \right] = {{\bf{u}}_e},\,\, \forall n \in {\cal N}, \tag{\ref{XX}{f}} \label{XXf}
\end{align}
where $\lambda $ is the slack variable. To simplify the mixed-integer non-convex problem, we first slacken the constraint \eqref{XXb} to $0 \le {x_m}\left[ w \right] \le 1,\forall m \in {\cal M},w \in {\cal W}$. Each time slot is divided into $Z$ time blocks. Then, for any given trajectory ${{\bf{u}}_1}\left[ w \right]$, the transmission scheduling ${\bf{X}}$ can be obtained by solving the following problem.
\begin{align}
	\textup{(sP1-1)}:\,\, &\mathop {\max }\limits_{\lambda ,{\bf{X}}} \lambda \label{ZZ}\\
	\mbox{s.t.}\quad 	
	&0 \le {x_m}\left[ w \right] \le 1,{\kern 1pt} \forall m \in {\cal M},w \in {\cal W}, \tag{\ref{ZZ}{a}} \label{ZZa}\\
	&\eqref{XXc},\eqref{XXd}. \nonumber
\end{align}

Problem \textup{(sP1-1)} is a standard linear programming (LP), which can be simply solved. For given ${\bf{X}}$, ${{\bf{u}}_1}\left[ w \right]$ is optimized to maximize the minimum amount of data. Specifically, the problem can be expressed as
\begin{align}
	\textup{(sP1-2)}:\,\, &\mathop {\max }\limits_{\lambda ,\left\{ {{{\bf{u}}_1}\left[ w \right]} \right\}} \lambda  \label{WW}\\
	\mbox{s.t.}\quad 	
	&\eqref{XXa},\eqref{XXd},\eqref{XXe},\eqref{XXf}. \nonumber
\end{align}

\begin{algorithm}[!t]
	\caption{Proposed PSO-based for sub-problem 2}
	\begin{algorithmic}[1]
		\STATE \textbf{Input:} Trajectory ${\left\{ {{{\bf{u}}_1}\left[ w \right]} \right\}}$ and altitude $H$ of UAV 1, number of sensors $M$, number of UAVs $N$, parameters of sensor position uncertainty $(h, {{\bf{\tilde s}}_m}$ and $r_u)$, maximum iterations $r_{\rm max}$, population size $N_p$ in the PSO-based algorithm.
		\STATE \textbf{Initialization:}
		\STATE Obtain the UAV 1 POPs ${\bf{I}}_1^*$ according to the parameters.
		\STATE Initialize the position vector  ${{\bf{P}}^0} = \left\{ {{\bf{P}}_1^0, \ldots ,{\bf{P}}_{{N_p}}^0} \right\}$ and velocity vector ${{\bf{V}}^0} = \left\{ {{\bf{V}}_1^0, \ldots ,{\bf{V}}_{{N_p}}^0} \right\}$  of the particle swarm. The position of each particle ${\bf{P}}_i^0$ is a set of distances and angles to determine the POPs of the auxiliary UAV.
		\STATE Calculate the fitness $f\left( {{\bf{P}}_i^0} \right)$ for each particle, whose reciprocal is the average positioning error $e_i^0$.
		\STATE Set the historical optimal position of each particle ${\bf{Pbes}}{{\bf{t}}_i} \leftarrow {\bf{P}}_i^0$.
		\STATE Obtain the global optimal position of particle swarm. Find ${i^*} = \mathop {\arg \max }\limits_{i \in \left\{ {1, \ldots ,{N_p}} \right\}} \left\{ {f\left( {{\bf{Pbes}}{{\bf{t}}_i}} \right)} \right\}$, and set  ${\bf{Gbes}}{{\bf{t}}^0} \leftarrow {\bf{Pbes}}{{\bf{t}}_{{i^*}}}$. 
		\STATE Set the optimal particle ${{\bf{P}}^*} \leftarrow {\bf{Gbes}}{{\bf{t}}^0}$ and corresponding fitness ${f^*}$.
		\STATE $r \leftarrow 0$.
		\STATE \textbf{Repeat:}.
		\STATE Update the velocity vector ${{\bf{V}}^{r + 1}} = \left\{ {{\bf{V}}_1^{r + 1}, \ldots ,{\bf{V}}_{{N_p}}^{r + 1}} \right\}$  and position vector ${{\bf{P}}^{r + 1}} = \left\{ {{\bf{P}}_1^{r + 1}, \ldots ,{\bf{P}}_{{N_p}}^{r + 1}} \right\}$  of the population.
		\STATE Calculate the fitness $f\left( {{\bf{P}}_i^{r + 1}} \right)$ for each particle.
		\STATE For each particle, if $f\left( {{\bf{P}}_i^{r + 1}} \right) > f\left( {{\bf{Pbes}}{{\bf{t}}_i}} \right)$: ${\bf{Pbes}}{{\bf{t}}_i} \leftarrow {\bf{P}}_i^{r + 1}$.
		\STATE Find ${i^*} = \mathop {\arg \max }\limits_{i \in \left\{ {1, \ldots ,{N_p}} \right\}} \left\{ {f\left( {{\bf{Pbes}}{{\bf{t}}_i}} \right)} \right\}$, and set ${\bf{Gbes}}{{\bf{t}}^{r + 1}} \leftarrow {\bf{Pbes}}{{\bf{t}}_{{i^*}}}$.
		\STATE If $f\left( {{\bf{Gbes}}{{\bf{t}}^{r + 1}}} \right) > f\left( {{{\bf{P}}^*}} \right)$: ${{\bf{P}}^*} \leftarrow {\bf{Gbes}}{{\bf{t}}^{r + 1}}$.
		\STATE \textbf{until} $r = {r_{\max }}$.
		\STATE Calculate auxiliary UAV POPs $\left\{ {{\bf{I}}_{_2}^*, \ldots {\bf{I}}_{_N}^*} \right\}$  according to ${{\bf{P}}^*}$.
		\STATE ${{\bf{I}}^*} = \left\{ {{\bf{I}}_{_1}^*, \ldots {\bf{I}}_{_N}^*} \right\}$ and minimum average positioning error ${e^*}$.
	\end{algorithmic}
\end{algorithm}

Due to constraints \eqref{XXa} and \eqref{XXd}, \textup{(sP1-2)} is still a non-convex optimization problem. To deal with the non-convex constraint \eqref{XXa}, the variable $\varsigma \left[ w \right] \ge 0$ is introduced. Let $\varsigma \left[ w \right] = {\left( {\sqrt {1 + \frac{{{{\left\| {{{\bf{v}}_1}\left[ w \right]} \right\|}^4}}}{{4v_0^4}}}  - \frac{{{{\left\| {{{\bf{v}}_1}\left[ w \right]} \right\|}^2}}}{{2v_0^2}}} \right)^{{1 \mathord{\left/
				{\vphantom {1 2}} \right.
				\kern-\nulldelimiterspace} 2}}}$, i.e., $\frac{1}{{\varsigma {{\left[ w \right]}^2}}} = \varsigma {\left[ w \right]^2} + \frac{{{{\left\| {{{\bf{v}}_1}\left[ w \right]} \right\|}^2}}}{{v_0^2}}$ \cite{8714077}. Then, problem \textup{(sP1-2)} is transformed into
\begin{align}
	\textup{(sP1-3)}:\,\, &\mathop {\max }\limits_{\lambda ,\left\{ {{{\bf{u}}_1}\left[ w \right]} \right\},\left\{ {\varsigma \left[ w \right]} \right\}} \lambda  \label{ZZ1}\\
	\mbox{s.t.}\quad 	
	&\tau \sum\limits_{w = 1}^W {P\left( {\left\| {{{\bf{v}}_1}\left[ w \right]} \right\|,\varsigma \left[ w \right]} \right)}  \le {E_{\max }}, \tag{\ref{ZZ1}{a}} \label{ZZ1a}\\
	&\varsigma {\left[ w \right]^2} + \frac{{{{\left\| {{{\bf{v}}_1}\left[ w \right]} \right\|}^2}}}{{v_0^2}} \ge {\kern 1pt} \frac{1}{{\varsigma {{\left[ w \right]}^2}}},\,\, \forall w \in {\cal W}{\kern 1pt}, \tag{\ref{ZZ1}{b}} \label{ZZ1b}\\
	&\eqref{XXd},\eqref{XXe},\eqref{XXf}. \nonumber
\end{align}
where
\begin{equation}
	\begin{aligned}
		P\left( {\left\| {{{\bf{v}}_1}\left[ w \right]} \right\|,\varsigma \left[ w \right]} \right)
		&= {k_b}\left( {1 + \frac{{3{{\left\| {{v_1}\left[ w \right]} \right\|}^2}}}{{v_t^2}}} \right) \\
		&+ \frac{1}{2}\rho s{d_r}A{\left\| {{v_1}\left[ w \right]} \right\|^3} + {k_i}\varsigma \left[ w \right],
		\label{equ 15}
	\end{aligned}
\end{equation}			
\begin{equation}
	{R_m}\left[ w \right] = B{\log _2}\left( {1 + \frac{{{c_0}}}{{{{\left( {{{\left\| {{{\bf{u}}_1}\left[ w \right] - {{\bf{s}}_m}} \right\|}^2} + {{\left( {H - {h_m}} \right)}^2}} \right)}^{\frac{\alpha }{2}}}}}} \right),
	\label{equ 16}
\end{equation}
where ${c_0} = {{{P_t}{\beta _0}} \mathord{\left/
		{\vphantom {{{P_t}{\beta _0}} {{\sigma ^2}}}} \right.
		\kern-\nulldelimiterspace} {{\sigma ^2}}}$. According to \cite{8714077}, problems \textup{(sP1-2)} and \textup{(sP1-3)} are equivalent. To obtain an effective approximate solution, we use SCA technique to transform the problem into a series of convex problems, and maximize the lower bound of \textup{(sP1-3)}. Let $R_m^l\left[ w \right]$ be the first-order Taylor expansion of ${R_m}\left[ w \right]$ on feasible point ${u^l} \buildrel \Delta \over = {\left\| {{{\bf{u}}_1}\left[ w \right] - {{\bf{s}}_m}} \right\|^2}$ in the $l$-th iteration. Then,
\begin{equation}
	\begin{aligned}
		{R_m}\left[ w \right] &\ge  R_m^l\left[ w \right] \buildrel \Delta \over = BA_1^l\left[ w \right] - BA_2^l\left[ w \right]{\left\| {{{\bf{u}}_1}\left[ w \right] - {{\bf{s}}_m}} \right\|^2} \\&+ BA_2^l\left[ w \right]{\left\| {{\bf{u}}_1^l\left[ w \right] - {{\bf{s}}_m}} \right\|^2},
		\label{equ 17}
	\end{aligned}
\end{equation}
where the formulas of $A_1^l\left[ w \right]$ and $A_2^l\left[ w \right]$ are shown in \eqref{equ 18} and \eqref{equ 19} respectively.
\begin{figure*}[!t]
	\normalsize
	\setcounter{mytempeqncnt}{\value{equation}}
	\setcounter{equation}{22}
	\begin{equation}
		\label{equ 18}
		A_1^l\left[ w \right] = {\log _2}\left( {1 + \frac{{{c_0}}}{{{{\left( {{{\left\| {{\bf{u}}_1^l\left[ w \right] - {{\bf{s}}_m}} \right\|}^2} + {{\left( {H - {h_m}} \right)}^2}} \right)}^{{\alpha  \mathord{\left/
									{\vphantom {\alpha  2}} \right.
									\kern-\nulldelimiterspace} 2}}}}}} \right)
	\end{equation}
	\begin{equation}
		\label{equ 19}
		A_2^l\left[ w \right] = \frac{{{{\alpha {c_0}{{\log }_2}e} \mathord{\left/
						{\vphantom {{\alpha {c_0}{{\log }_2}e} 2}} \right.
						\kern-\nulldelimiterspace} 2}}}{{\left( {{{\left\| {{\bf{u}}_1^l\left[ w \right] - {{\bf{s}}_m}} \right\|}^2} + {{\left( {H - {h_m}} \right)}^2}} \right)\left( {{{\left( {{{\left\| {{\bf{u}}_1^l\left[ w \right] - {{\bf{s}}_m}} \right\|}^2} + {{\left( {H - {h_m}} \right)}^2}} \right)}^{{\alpha  \mathord{\left/
									{\vphantom {\alpha  2}} \right.
									\kern-\nulldelimiterspace} 2}}} + {c_0}} \right)}}
	\end{equation}
	\setcounter{equation}{\value{mytempeqncnt}}
	\hrulefill
	\vspace*{4pt}
\end{figure*}
Similarly, for the first-order Taylor expansion of \eqref{ZZ1b} at ${\varsigma ^l}\left[ w \right]$ and ${\bf{v}}_1^l\left[ w \right]$, 
\setcounter{equation}{24} 
\begin{equation}
	\begin{aligned}
		\varsigma {\left[ w \right]^2} &+ \frac{{{{\left\| {{{\bf{v}}_1}\left[ w \right]} \right\|}^2}}}{{v_0^2}} \ge {\vartheta ^l}\left[ w \right] \buildrel \Delta \over =  - {\varsigma ^l}{\left[ w \right]^2} + 2{\varsigma ^l}\left[ w \right]\varsigma \left[ w \right] \\
		&- \frac{{{{\left\| {{\bf{v}}_1^l\left[ w \right]} \right\|}^2}}}{{v_0^2}} + \frac{2}{{v_0^2}}\left\| {{\bf{v}}_1^l\left[ w \right]} \right\| \cdot \left\| {{{\bf{v}}_1}\left[ w \right]} \right\|.
		\label{equ 20}
	\end{aligned}
\end{equation}

By introducing the lower bound $R_m^l\left[ w \right]$ and ${\vartheta ^l}\left[ w \right]$, problem \textup{(sP1-3)} can be approximated as follows in the $l$-th iteration.
\begin{align}
	\textup{(sP1-4)}:\,\, &\mathop {\max }\limits_{\lambda ,\left\{ {{{\bf{u}}_1}\left[ w \right]} \right\},\left\{ {\varsigma \left[ w \right]} \right\}} \lambda  \label{P6}\\
	\mbox{s.t.}\quad 	
	&{\vartheta ^l}\left[ w \right] \ge {\kern 1pt} \frac{1}{{\varsigma {{\left[ w \right]}^2}}},\,\, \forall w \in {\cal W}, \tag{\ref{P6}{a}} \label{P6a}\\
	&\frac{\tau }{{{I_m}}}\sum\limits_{w = 1}^W {R_m^l\left[ w \right] \cdot {x_m}\left[ w \right]}  \ge {\kern 1pt} {\kern 1pt} \lambda ,\,\, \forall m \in {\cal M}, \tag{\ref{P6}{b}} \label{P6b}\\
	&\eqref{ZZ1a},\eqref{XXe},\eqref{XXf}. \nonumber
\end{align}

Problem \textup{(sP1-4)} is a convex optimization problem, which can be solved by CVX. To sum up, by solving problems \textup{(sP1-1)} and \textup{(sP1-4)} alternately, we can obtain a local optimal solution of \textup{(sP1)}.

\begin{figure}[t]
	\centering
	\includegraphics[width=0.37\textheight]{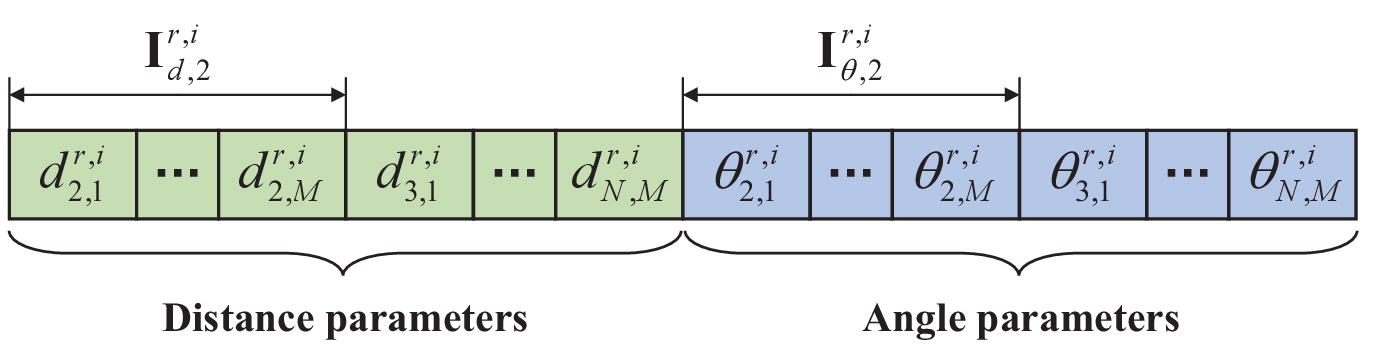}
	\DeclareGraphicsExtensions.
	\caption{ Composition of the $i$-th particle ${\bf{P}}_{_i}^r$ in the $r$-th iteration.}
	\label{fig 4}
\end{figure}

\subsection{\textup{(sP2)}: PoPs and Auxiliary UAV Trajectory Optimization}
Based on the optimal UAV 1 trajectory obtained in \textup{(sP1)},
we can find the trajectories of auxiliary UAVs that meet the
positioning requirements. However, the objective function of
\textup{(sP2)} does not have a closed-form expression, which is difficult to be
solved by conventional optimization methods. To simplify the
problem and obtain a feasible solution, we propose a PSO-based algorithm to optimize POPs of UAVs. The detailed
process is shown in Algorithm 2, and the obtained solution is local optimal. To select the POPs of UAV
1, we first explore the effect of the distance between UAV
1 and the sensor on the positioning performance. As shown
in \textcolor[rgb]{0,0.4471,0.4039}{Fig. \ref{fig 3}}, the RMSE of the sensor almost linearly increases
with the horizontal distance changes from 10 m to 100 m.
Therefore, the UAV 1 position nearest to the sensor is selected
as the POP.

In the PSO algorithm, particles have two attributes: position
and velocity. The position is considered as a candidate solution
of the optimization problem. In a iteration, the particle tracks
the extreme values of its individual and entire population in
the search space and adjusts its position and velocity to find
a satisfactory solution. Since the auxiliary UAVs needs to be
within the communication range of UAV 1 during positioning,
we define the position vector of each particle as the set of the
distance and angle between the auxiliary UAV and UAV 1,
as shown in \textcolor[rgb]{0,0.4471,0.4039}{Fig. \ref{fig 4}}. Among them, ${\bf{I}}_{d,2}^{r,i}$ and ${\bf{I}}_{\theta ,2}^{r,i}$  represent the distance and angle of the auxiliary UAV $2$ relative to UAV $1$ at the POP, respectively. The distance parameter meets the constraint of communication range, namely $d_{n,m}^{r,i} \le {R_{\max }}$.

\begin{figure}[!htbp]
	\centering
	\hspace{-30pt}
	\subfigure[]{
		\label{fig 5(a)}
		\includegraphics[width=0.4\textwidth]{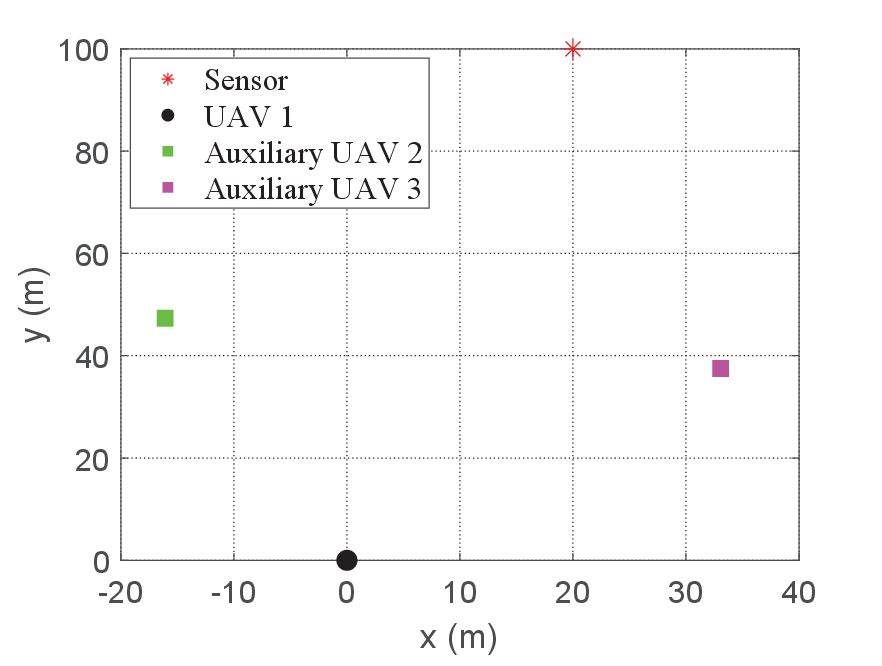}}
	
	\hspace{-30pt}
	\subfigure[]{
		\label{fig 5(b)}
		\includegraphics[width=0.4\textwidth]{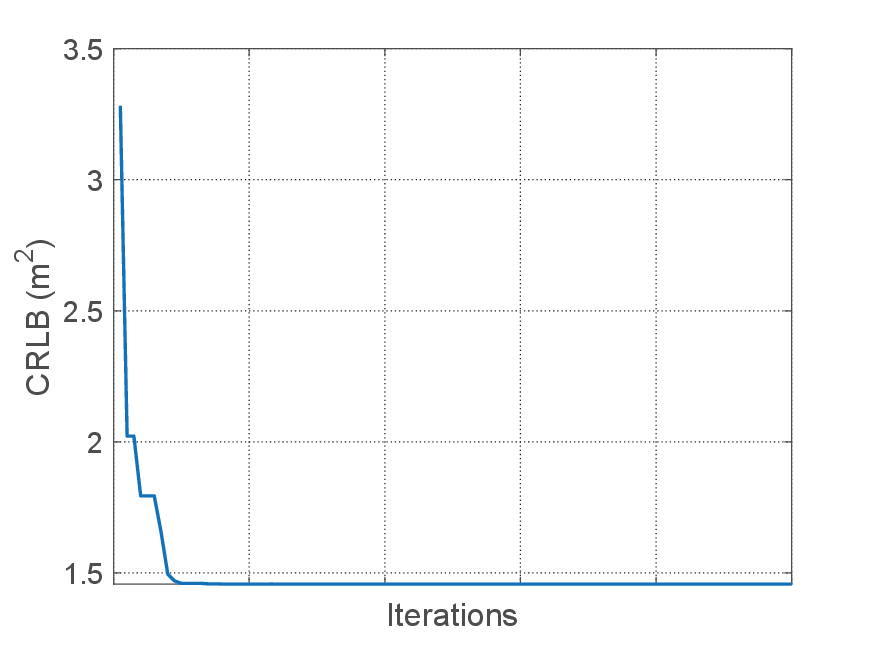}}
	
	\caption{POPs Optimization using the proposed PSO-based algorithm with $N = 3$ and ${R_{\max }} = 50$ m. (a)The optimal POPs. (b)Corresponding iterative curve.}
	\label{fig 5}
\end{figure}

\begin{figure*}[t]
	\begin{center}
		\hspace{-34pt}
		\subfigure[]{
			\label{fig 6(a)}
			\includegraphics[width=2.5in,height=2.1in]{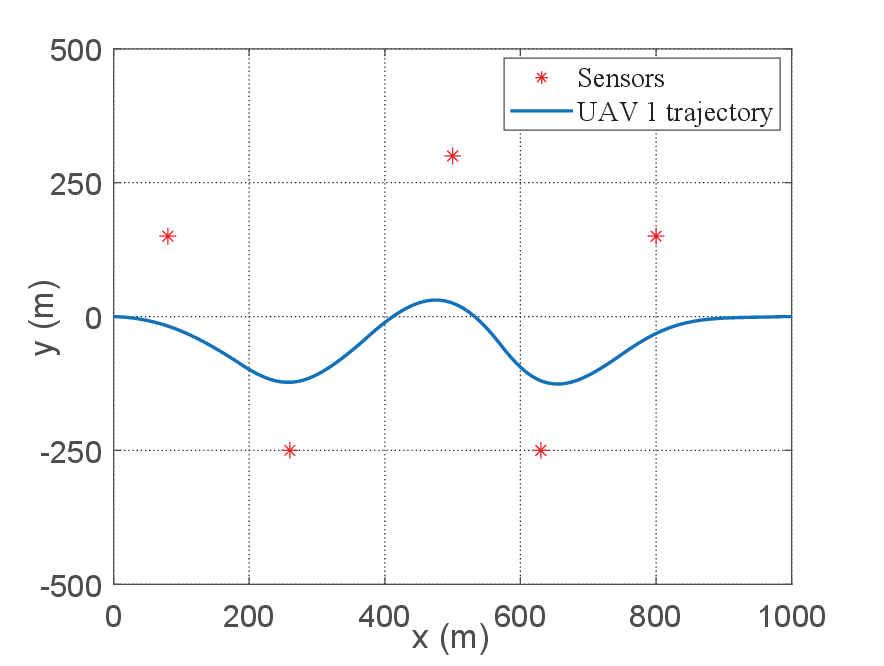}
			\hspace{-25pt}
A		}
		\subfigure[]{
			\label{fig 6(b)}
			\includegraphics[width=2.5in,height=2.1in]{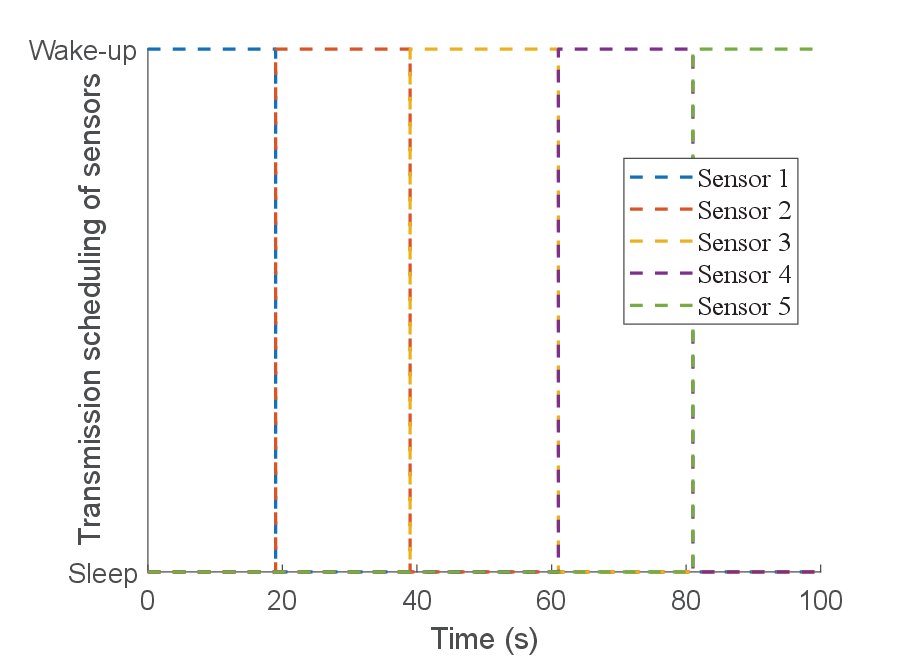}
			\hspace{-25pt}
		}
		\subfigure[]{
			\label{fig 6(c)}
			\includegraphics[width=2.5in,height=2.1in]{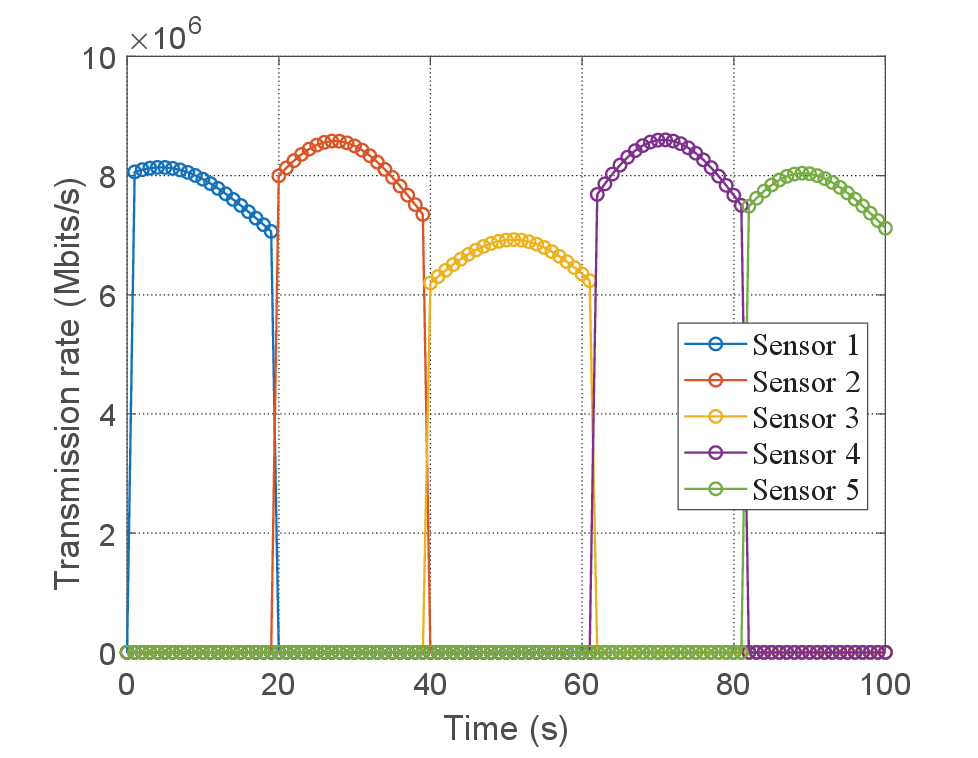}
		}
		\hspace{-40pt}
		\caption{Optimized (a) UAV 1 trajectory, (b) sensor transmission scheduling and (c) data transmission rate in scenario 1. }
		\label{fig 6}
	\end{center}
\end{figure*}

\begin{figure*}[t]
	\begin{center}
		\hspace{-34pt}
		\subfigure[]{
			\label{fig 7(a)}
			\includegraphics[width=2.5in,height=2.1in]{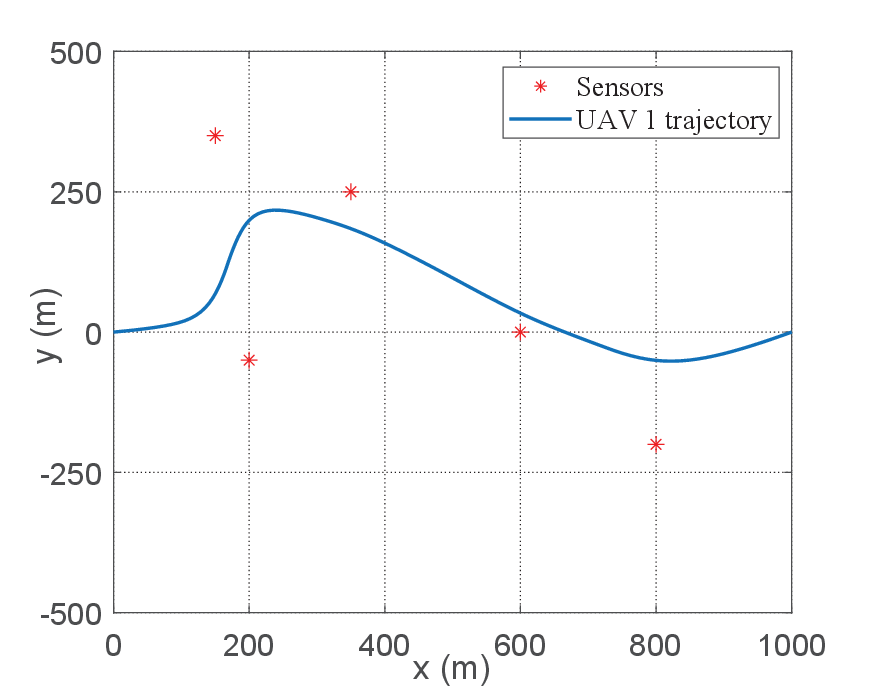}
			\hspace{-25pt}
A		}
		\subfigure[]{
			\label{fig 7(b)}
			\includegraphics[width=2.5in,height=2.1in]{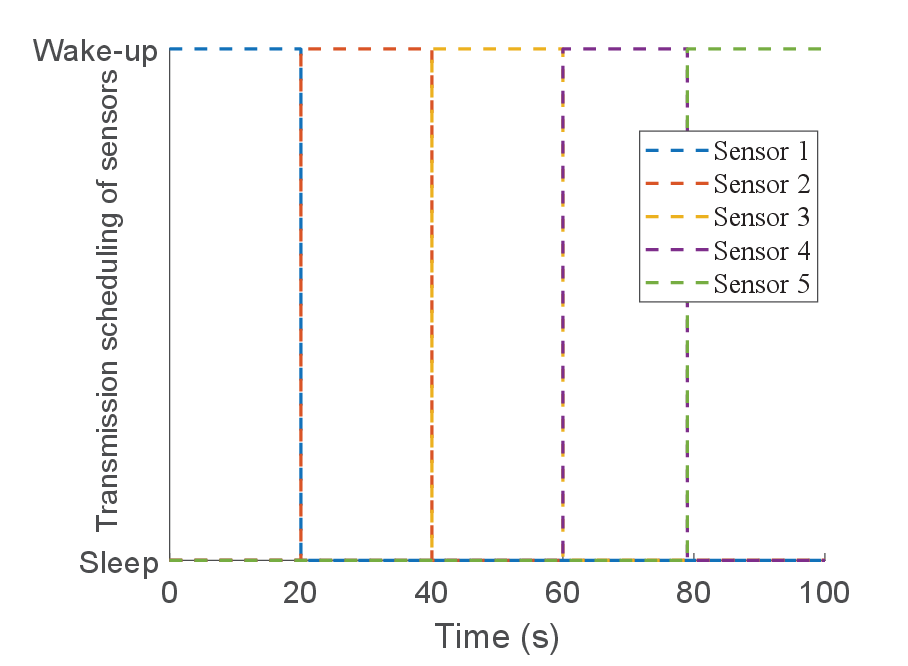}
			\hspace{-25pt}
		}
		\subfigure[]{
			\label{fig 7(c)}
			\includegraphics[width=2.5in,height=2.1in]{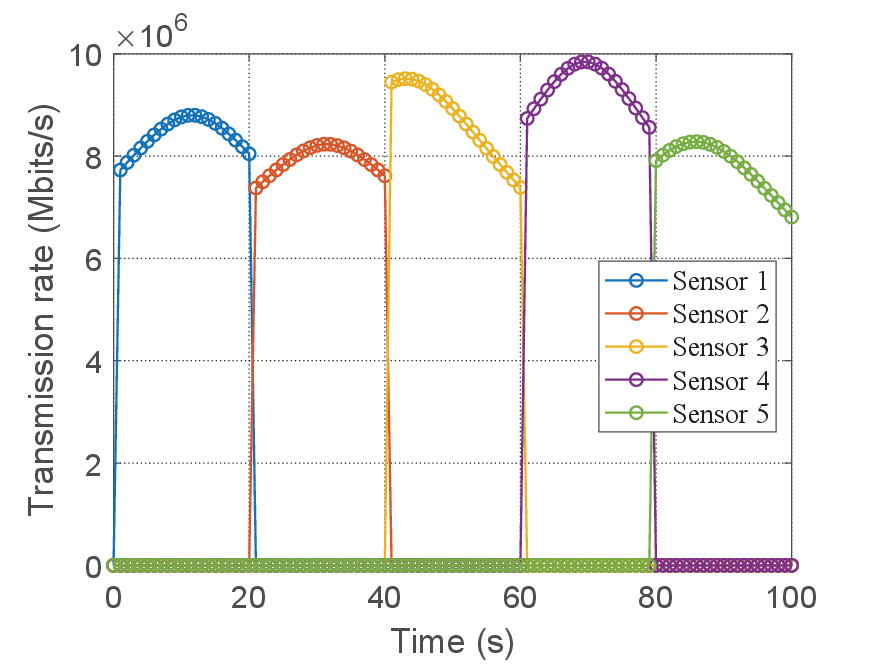}
		}
		\hspace{-40pt}
		\caption{Optimized (a) UAV 1 trajectory, (b) sensor transmission scheduling and (c) data transmission rate in scenario 2.}
		\label{fig 7}
	\end{center}
	
\end{figure*}

To minimize the average positioning error of $M$ sensors while satisfying the constraints \eqref{YYa} and \eqref{YYe}, we set the following fitness function \cite{8894454}.
\begin{equation}
	fit\left( {{\bf{P}}_{_i}^l} \right) = \frac{1}{{{c_{\rm{CRLB}}}{f_{\rm{CRLB}}} + {c_e}{f_e} + {c_v}{f_v}}},
	\label{equ 21}
\end{equation} 
where $f_e$ and $f_v$ are binary variables and represent the penalty items of energy consumption and speed constraints. $f_{\rm{CRLB}}$ is the average CRLB of all sensors. When the above constraint is violated, let ${f_e} = 0$ or ${f_v} = 0$. $c_{\rm{CRLB}}$, $c_{e}$ and $c_{v}$ are the corresponding penalty factors, and the values of $c_{e}$ and $c_{v}$ should be much greater than the values of $c_{\rm{CRLB}}$, so as to give the infeasible particle a lower fitness value. \eqref{equ 21} ensures that the optimal solution obtained in each iteration is feasible.

\begin{table}[!t]
	\caption{Simulation parameters}
	\label{table}
	\setlength{\tabcolsep}{3pt}
	\begin{tabular}{|m{6.7cm}<{\centering}|m{2cm}<{\centering}|}
		\hline
		\textbf{Parameter}& \textbf{Value} \\
		\hline
		Length of target area ($L$) & $1$ km \\
		\hline
		Number of UAVs ($N$)& $3$ \\
		\hline
		Number of sensors ($M$)& $5$ \\
		\hline
		UAV flight altitude ($H$)& $100$ m \\
		\hline
		Sensor height ($h_m$)& $0$ m \\
		\hline
		Number of time slots ($W$)& $100$ \\
		\hline
		Length of time slots ($\tau $)& $1$ s \\
		\hline
		Carrier frequency ($f$)& $2.1$ GHz\\
		\hline
		Channel bandwidth ($B_0$)& $1$ MHz \\
		\hline
		Reference channel power gain at $1$ m (${\beta _0}$) & $-60$ dB\\
		\hline
		Path loss factor ($\alpha$)& $2$\\
		\hline
		Noise power ($\delta ^2$)&  $-110$ dBm\\
		\hline
		Transmit power of sensors ($P_t$) & $0.1$ W\\
		\hline
		Maximum speed of UAVs (${V_{\max }}$) & $30$ m/s \\
		\hline
		Maximum communication range between UAVs (${R_{\max }}$)& $50$ m \\
		\hline
		Maximum number of sensors communicating with UAV ($K_{\max}$) & $1$\\
		\hline
		Radius of uncertainty location area ($r_u$)& $10$ m\\
		\hline
		amount of data uploaded by each sensor (${I_{\min }}$) & $20$ Mbits\\
		\hline
		Initial position of UAVs (${{\bf{u}}_s}$) & ${\left( {0,0} \right)^T}$ m\\
		\hline
		Final position of UAVs (${{\bf{u}}_e}$) & ${\left( {1000,0} \right)^T}$ m\\
		\hline
		Population size ($N_p$) & $50$\\
		\hline

		Maximum iterations of BCD algorithm  (${l_{\max }}$) & $30$\\
		\hline
		Maximum iterations of PSO algorithm  (${r_{\max }}$)& $200$\\
		\hline
	\end{tabular}
	\label{tab1}
\end{table}

To balance the global search ability and local search ability
of the algorithm, we adopt the strategy of inertia weight
linearly decreasing, as shown in \eqref{equ 22}.
\begin{equation}
	{\omega ^r} = {\omega _{\max }} - \frac{{\left( {{\omega _{\max }} - {\omega _{\min }}} \right)r}}{{{r_{\max }}}},
	\label{equ 22}
\end{equation} 
where ${\omega ^r}$ is the linearly decreasing adaptive inertia weight, and $r$ is the current iteration number. ${\omega _{\max }}$ and ${\omega _{\min }}$ are the final and initial values of inertia weight respectively, with typical values of $0.9$ and $0.4$ \cite{785511}. With the increasing of iterations,  ${\omega ^r}$ decreases gradually, and the global search ability and local search ability of PSO are weakened and enhanced respectively. The particle updates itself by tracking the optimal solution found by itself (individual extremum ${\bf{Pbes}}{{\bf{t}}_i}$) and the optimal solution found by the whole population (global extremum ${\bf{Gbes}}{{\bf{t}}^r}$). Each particle updates its speed and position according to \eqref{equ 23} and \eqref{equ 24}.
\begin{equation}
	\begin{aligned}
		{\bf{V}}_i^{r + 1} &= \underbrace {{\omega ^r}{\bf{V}}_i^r}_{\;{\textup{Memory item}}} + \underbrace {{c_1} \cdot rand\left( {} \right) \cdot \left( {{\bf{Pbes}}{{\bf{t}}_i} - {\bf{P}}_i^r} \right)}_{{\textup{Self - cognition item}}}\\
		&+ \underbrace {{c_2} \cdot rand\left( {} \right) \cdot \left( {{\bf{Gbes}}{{\bf{t}}^r} - {\bf{P}}_i^r} \right)}_{{\textup{Group - cognition item}}},
		\label{equ 23}
	\end{aligned}
\end{equation} 
\begin{equation}
	{\bf{P}}_i^{r + 1} = {\bf{P}}_i^r + {\bf{V}}_i^{r + 1},
	\label{equ 24}
\end{equation}
where $c_1$ and $c_2$ are learning factors,  $rand\left( {} \right)$  is a random number between 0 and 1, and ${\bf{V}}_i^r$  and  ${\bf{P}}_i^r$ are the velocity and position of the $i$-th particle with certain limits. When the values ${\bf{V}}_i^r$  and  ${\bf{P}}_i^r$ are outside the feasible range, they will be assigned to the minimum or maximum values within the feasible range.

To verify the feasibility of the proposed algorithm, a simple scenario is designed for testing. We set ${{\bf{u}}_1} = \left[ {0,0} \right]$  and ${\bf{s}} = \left[ {20,100} \right]$, and the optimal POPs of UAVs is shown in \textcolor[rgb]{0,0.4471,0.4039}{Fig. \ref{fig 5(a)}}. \textcolor[rgb]{0,0.4471,0.4039}{Fig. \ref{fig 5(b)}} shows the CRLB decreases significantly with the number of iterations increases.

\section{Numerical Results}
In this section, a series of numerical simulation results
are provided to verify the proposed scheme and evaluate its
performance. The key simulation parameters are summarized in \textcolor[rgb]{0,0.4471,0.4039}{Table \ref{tab1}} \cite{8714077,9013148,8894454,8647595}.

\begin{figure*}[htbp]
	\begin{minipage}[t]{0.45\linewidth}
		\centering
		\includegraphics[width=0.36\textheight]{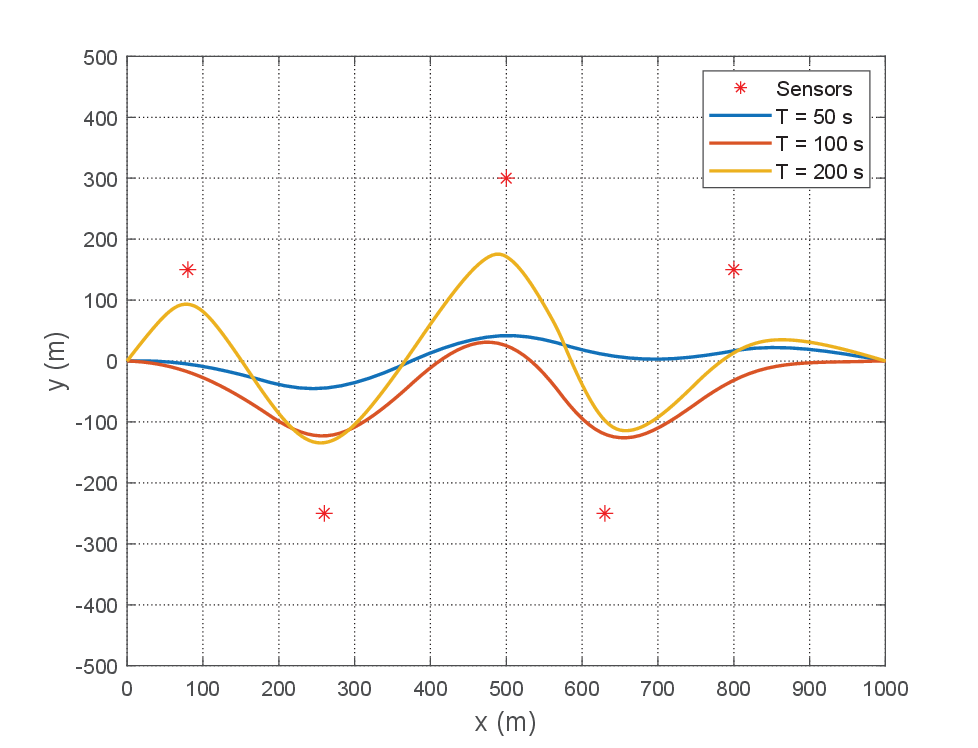}
		\caption{ Optimized UAV $1$ trajectories with different $T$.}
		\label{fig 8}	
	\end{minipage}%
	\hspace{8mm}
	\begin{minipage}[t]{0.45\linewidth}
		\centering
		\includegraphics[width=0.36\textheight]{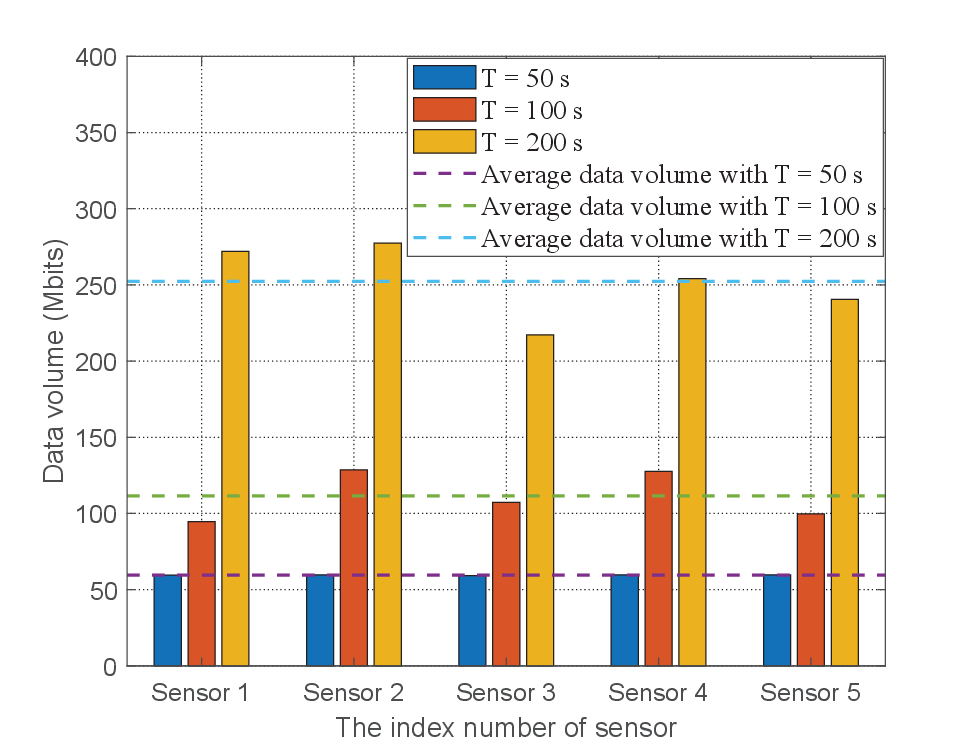}
		\caption{The amount of data uploaded by each sensor and the average data amount.}
		\label{fig 9}
	\end{minipage}
\end{figure*}

\begin{figure*}[htbp]
	\begin{minipage}[t]{0.45\linewidth}
		\centering
		\includegraphics[width=8.8cm,height=7cm]{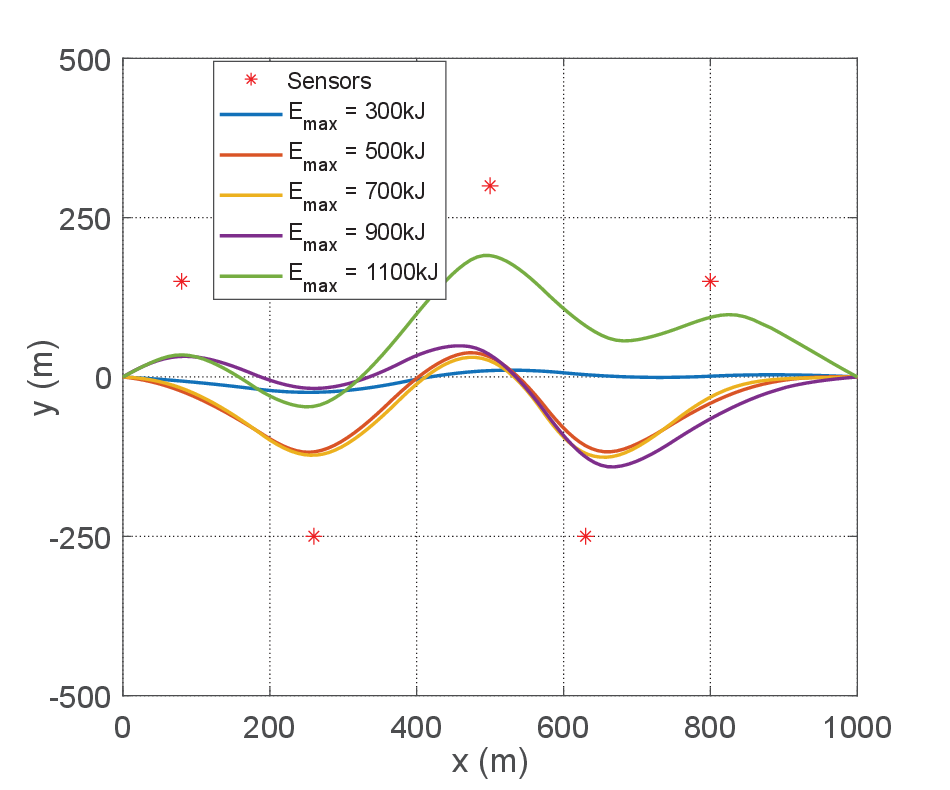}
		\caption{Optimized UAV $1$ trajectories for different $E_{\max}$.}
		\label{fig 10}	
	\end{minipage}%
	\hspace{8mm}
	\begin{minipage}[t]{0.45\linewidth}
		\centering
		\includegraphics[width=8.8cm,height=7.1cm]{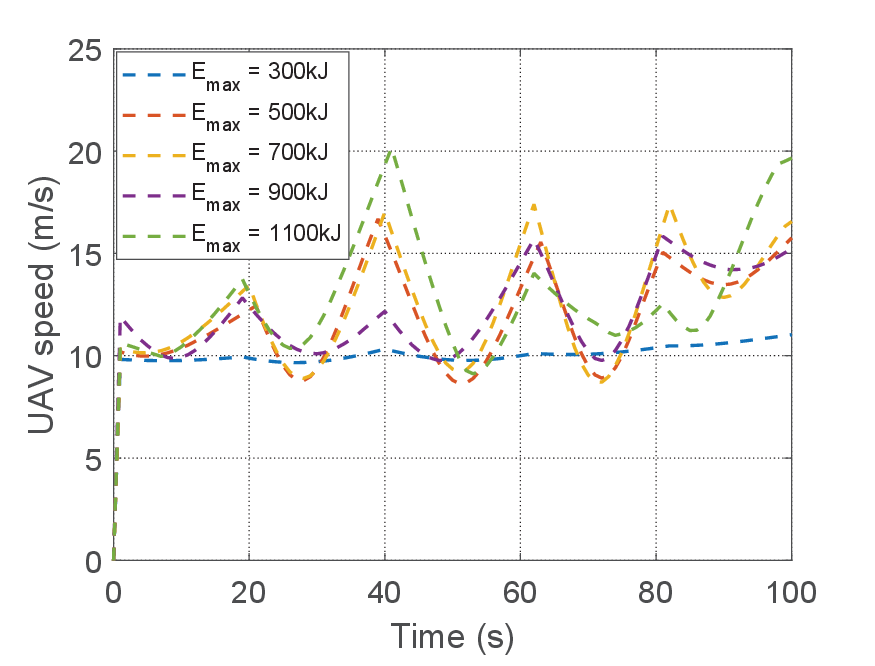}
		\caption{ Optimized UAV $1$ speed for different $E_{\max}$.}
		\label{fig 11}
	\end{minipage}
\end{figure*}

\subsection{Performance Evaluation of Data Collection}

\textcolor[rgb]{0,0.4471,0.4039}{Fig. \ref{fig 6}} and \textcolor[rgb]{0,0.4471,0.4039}{Fig. \ref{fig 7}} show the optimized trajectory of UAV 1, sensor transmission scheduling and instantaneous data transmission rate with $E_{\rm max} = 700$ KJ under the two sensor deployment scenarios. Obviously, the trajectory of UAV 1 in \textcolor[rgb]{0,0.4471,0.4039}{Fig. \ref{fig 6(a)}} and \textcolor[rgb]{0,0.4471,0.4039}{Fig. \ref{fig 7(a)}} tends to be close to the sensor, and the sensor with relatively close distance is selected for data collection. This is due to the fact that the channel quality is high when the
sensor is close. The above conclusion is also confirmed by the
change of data transmission rate in \textcolor[rgb]{0,0.4471,0.4039}{Fig. \ref{fig 6(b)}} and \textcolor[rgb]{0,0.4471,0.4039}{Fig. \ref{fig 7(b)}}. \textcolor[rgb]{0,0.4471,0.4039}{Fig. \ref{fig 6(c)}} and \textcolor[rgb]{0,0.4471,0.4039}{Fig. \ref{fig 7(c)}} show that the sensor will remain in sleep state until UAV 1 approaches it and wakes it up, and the
communication time with each sensor is roughly the same.
In addition, in order to maximize the minimum amount of
data uploaded by the sensor, the UAV 1 will always maintain 
a communication link with the sensor without restricting the
energy consumption of the sensor.

\textcolor[rgb]{0,0.4471,0.4039}{Fig. \ref{fig 8}} shows the  optimized trajectory of UAV 1 in the horizontal direction when the mission cycle $T$ is 50 s, 100 s and 200 s respectively. As expected, with the increase of $T$, UAV 1 tends to be closer to sensors to improve the efficiency of data collection. The amount of data uploaded by each sensor and the average amount of data with different $T$ are shown in \textcolor[rgb]{0,0.4471,0.4039}{Fig. \ref{fig 9}}. Through the trend of the total data amount, it can also be found that the sensors close to the flight trajectory can always upload more data.

\textcolor[rgb]{0,0.4471,0.4039}{Fig. \ref{fig 10}} and \textcolor[rgb]{0,0.4471,0.4039}{Fig. \ref{fig 11}} show the impact of energy consumption
constraint on the trajectory and speed of UAVs. Given $T = 100$ s, as available energy increase, UAV 1 tends to fly at a higher rate to obtain a longer trajectory, which is consistent with the expected results. It is noted that the trajectories under $E_{\max} = 500$ KJ and $E_{\max} = 700$ KJ are very similar. \textcolor[rgb]{0,0.4471,0.4039}{Fig. \ref{fig 11}} also shows that the trends of speed change in both cases is also roughly the same. When $E_{\max} = 700$ KJ, the average rate is higher, resulting in higher energy consumption.

\subsection{Performance Evaluation of Sensor Positioning}

\begin{figure*}[!t]
	\begin{center}
		\hspace{-34pt}
		\subfigure[]{
			\label{fig 12}
			\includegraphics[width=2.6in,height=2.15in]{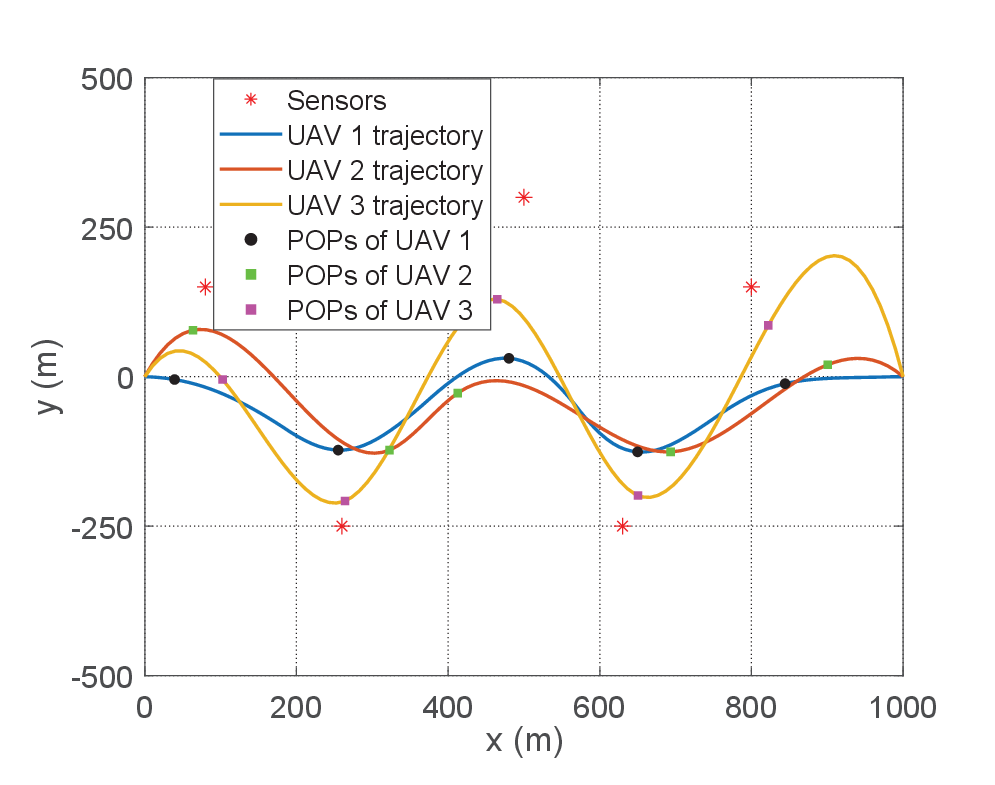}
			\hspace{-25pt}
		}
		\subfigure[]{
			\label{fig 13}
			\includegraphics[width=2.6in,height=2.1in]{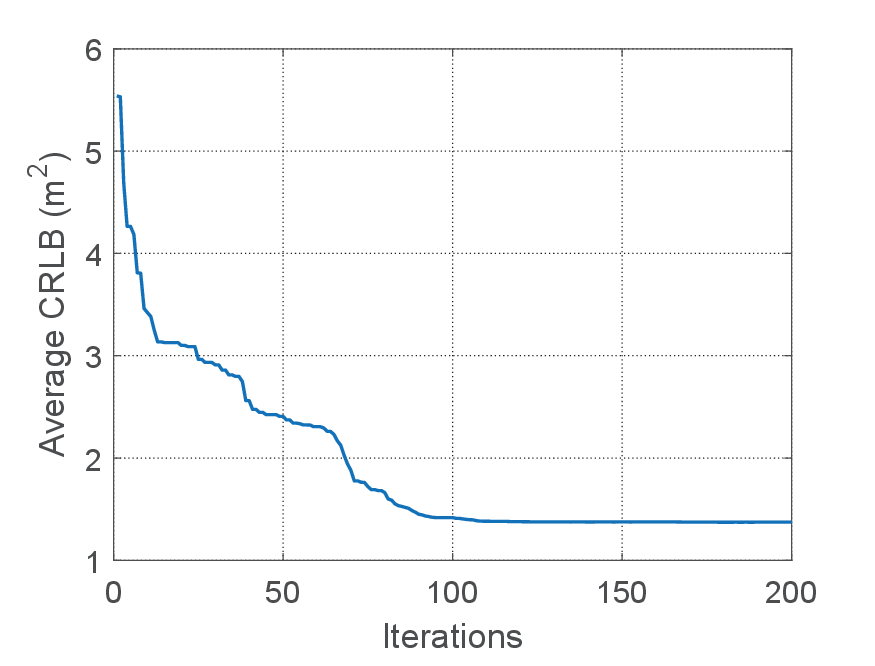}
			\hspace{-25pt}
		}
		\subfigure[]{
			\label{fig 14}
			\includegraphics[width=2.6in,height=2.1in]{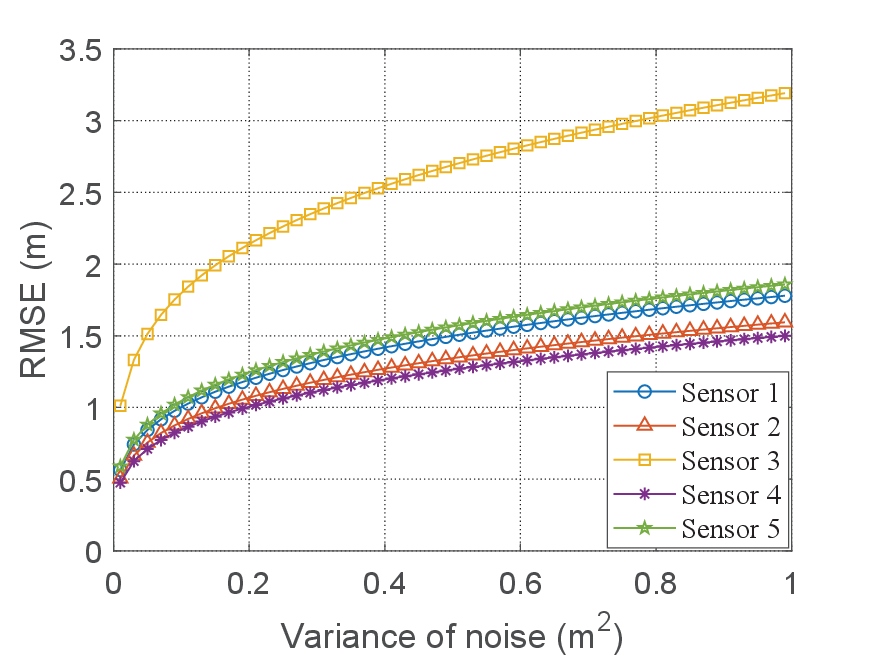}
		}
		\hspace{-40pt}
		\caption{Application results of the proposed PSO-based algorithm. (a) Optimized UAV trajectories and POPs. (b) Corresponding iterative curve. (c) Influence of noise on positioning accuracy. }
		\label{fig 15}
	\end{center}
\end{figure*}

\textcolor[rgb]{0,0.4471,0.4039}{Fig. \ref{fig 12}} shows the POPs and trajectories of auxiliary UAV obtained by Algorithm 2 based on the optimized trajectory of UAV 1 with $T = 100$ s and $E_{\max} = 700$ KJ. It is shown that in order to improve the positioning accuracy, UAVs are distributed as far as possible within the maximum communication range, which is consistent with the assumption. As shown in \textcolor[rgb]{0,0.4471,0.4039}{Fig. \ref{fig 13}}, as the number of iterations increases, the fitness of the optimal individual of the particle population continues to improve, and the average positioning error of the sensor decreases significantly.

The positioning at the POPs that are obtained in \textcolor[rgb]{0,0.4471,0.4039}{Fig. \ref{fig 12}}
is executed to study the relation between the positioning error and the variance of noise. As shown in \textcolor[rgb]{0,0.4471,0.4039}{Fig. \ref{fig 14}}, with the
increase of the variance of noise, the positioning error of
sensor shows a logarithmic growth trend. When the sensor is far from the anchor node, its positioning error is significantly
higher than other sensors. This phenomenon inspires us to
make the UAV 1 close to the sensor, so as to improve
the positioning accuracy by increasing the signal-noise ratio
(SNR) threshold.

\section{Conclusion}
This paper proposes a joint data collection and sensor
positioning scheme for WSN using multiple UAVs. Firstly, the
CRLB of sensor positioning with TDoA is derived to evaluate
positioning performance. Then, a mixed-integer non-convex
optimization model is established considering the constraints
of the flight energy consumption, flight speed, requirements
for data collection and communication range. The optimization
model is to optimize the UAV flight trajectory, sensor transmission schedule and POPs jointly to minimize the average
positioning error of sensors, while ensuring the amount of
data collection. Then, BCD, SCA and PSO-based optimization
algorithms are applied to solve the optimization model. The
numerical simulation results show that the proposed scheme
achieves efficient and reliable data collection, and have a good
positioning performance. This paper may motivate the joint
design of data collection and sensor positioning for WSN
deployed in the areas that lack infrastructure coverage.


\bibliographystyle{IEEEtran}
\bibliography{literatures}

\end{document}